\documentclass[fleqn,usenatbib]{mnras}

\usepackage{newtxtext,newtxmath}

\usepackage[T1]{fontenc}
\usepackage{ae,aecompl}

\usepackage{graphicx}  
\usepackage{dcolumn}   
\usepackage{bm}        
\usepackage{comment}
\usepackage{amssymb}   
\usepackage{amsmath}
\usepackage{tabularx}
\usepackage{hyperref}
\usepackage{tabulary}
\usepackage[usenames,dvipsnames]{color}
\usepackage[normalem]{ulem}

\newcommand{\msun}{M_{\odot}}

\newcommand{\mspc}{M_{\odot}\,{\rm pc}^{-3}}

\newcommand{\be}{\begin{equation}}
\newcommand{\ee}{\end{equation}}
\newcommand{\bi}{\begin{list}{\labelitemi}{\leftmargin=1em}\setlength{\itemsep}{-3pt}}
\newcommand{\ei}{\end{list}}

\newcommand{\beqn}{\begin{eqnarray}}
\newcommand{\eeqn}{\end{eqnarray}}

\newcommand{\trh}{\tau_{\rm rh}}
\newcommand{\trhn}{\tau_{\rm rh,0}}
\newcommand{\rh}{r_{\rm h}}
\newcommand{\rhn}{r_{\rm h,0}}
\newcommand{\rhdot}{\dot{r}_{\rm h}}
\newcommand{\rhoh}{\rho_{\rm h}}
\newcommand{\mmean}{\langle m_\star\rangle}
\newcommand{\vesc}{v_{\rm esc}}
\newcommand{\vescn}{v_{\rm esc,0}}

\newcommand{\kms}{{\rm km\,s}^{-1}}
\newcommand{\myr}{{\rm Myr}}
\newcommand{\pc}{{\rm pc}}

\title[Black hole growth in star clusters]{Black hole growth  through hierarchical 
black hole mergers in dense star clusters:
implications for gravitational wave detections
}

\author[Antonini et al.]{
Fabio Antonini,$^{1}$\thanks{f.antonini@surrey.ac.uk}
Mark Gieles,$^{1,2,3}$
and
Alessia Gualandris,$^{1}$
\\
$^1$Faculty of Engineering and Physical Sciences, University of Surrey, Guildford, Surrey, GU2 7XH,
United Kingdom \\
$^2$Institut de Ci\`{e}ncies del Cosmos (ICCUB), Universitat de Barcelona, Mart\'{i} i Franqu\`{e}s 1, E08028 Barcelona, Spain\\
$^3$ICREA, Pg. Lluis Companys 23, 08010 Barcelona, Spain.
}

\date{Accepted XXX. Received YYY; in original form ZZZ}

\pubyear{2018}

\begin{document}
\label{firstpage}
\pagerange{\pageref{firstpage}--\pageref{lastpage}}
\maketitle

\begin{abstract}
In a  star  cluster  with a sufficiently  large  escape  velocity, black holes (BHs) that are produced by BH mergers can be retained, dynamically form  new {BH} binaries,  and  merge  again. This process can
repeat several times and lead to significant mass growth.
  In this paper, we calculate the  mass  of the largest BH that can form through repeated BH mergers and determine  how its value depends on the physical properties of the host cluster. We adopt an analytical model in which the energy generated by the black hole binaries in the cluster core is assumed to be regulated by the process of two-body relaxation in the bulk of the system. This  principle is used to compute the hardening rate of the binaries and to relate this to the time-dependent global properties of the parent cluster. 
We demonstrate that in clusters with initial escape velocity  $\gtrsim 300\, \kms$ in the core and  density
$\gtrsim 10^5\, M_\odot\rm pc^{-3}$,  repeated mergers lead to the formation of BHs  in the mass range $100-10^5 \,M_\odot$, populating any upper mass gap created by pair-instability supernovae.
 This result is independent of cluster metallicity  and the initial BH spin distribution. We show that about $10\%$ of the present-day nuclear star clusters meet these extreme conditions, and  estimate that BH binary mergers with total mass $\gtrsim 100\,M _\odot$ should be produced in these systems at a maximum rate $\approx 0.05 \,\rm Gpc^{-3} yr^{-1}$, corresponding to  one detectable event every few years with Advanced LIGO/Virgo at design sensitivity.\end{abstract}




\section{introduction}
Dynamical three-body  interactions 
 in the dense core of a star cluster
 lead to the efficient formation, hardening and merger
 of black hole (BH) binaries \citep{Kulkarni1993,Sigurdsson1993}.
Numerical simulations based on Monte Carlo and $N$-body techniques  \citep[e.g.,][]{Zwart1999,Downing2010,Banerjee2010,2013MNRAS.435.1358T,Rodriguez2015a,Rodriguez2016a,2017MNRAS.464L..36A,2018ApJ...855..124S} have shown 
 that such  dynamically formed binaries could explain many, or perhaps even
 most, of the BH mergers that have been observed by Advanced LIGO and Virgo \citep{2016PhRvL.116f1102A,2017PhRvL.119n1101A}. 
However, it remains an open question  what physical properties will
allow to distinguish these  systems
from  binaries which are produced  in the field of a galaxy   through either the  evolution of isolated pairs of massive stars \citep{2002ApJ...572..407B,2016MNRAS.458.3075A,Belczynski2016,Marchant2016,Mandel2016a,2018arXiv180802491G,2018arXiv180904605S,2018PhRvD..98f3018A}
or in hierarchical triple systems via the Lidov-Kozai mechanism
\citep{2017ApJ...836...39S,2017ApJ...841...77A,2018MNRAS.480L..58A,2018ApJ...863....7R,2018ApJ...863...68L}. Currently, both 
 cluster and field formation scenarios
appear to be broadly consistent with the Advanced LIGO/Virgo 
detections \citep{2016ApJ...818L..22A}.

A possible way to determine which formation mechanism is dominant is to 
compare the mass function of the detected binaries to model predictions.
The BHs
that are produced through three body dynamical processes in a dense star cluster might in fact
be
much heavier than those formed from 
the collapse of massive stars.
Field formation scenarios  predict long  times between the formation and merger of a compact binary \citep[e.g.,][]{2002ApJ...572..407B}, so that repeated  mergers should be nearly impossible in this case.
On the other hand, in star clusters with sufficiently large escape velocities, 
 BHs that are produced by previous mergers
can be retained 
after the momentum kick due to the merger, form new binaries, and merge again  \citep{2016ApJ...831..187A,2018PhRvL.120o1101R}.
Statistically observable imprints in the properties of the newly formed BHs are expected \citep{2016ApJ...831..187A,2017PhRvD..95l4046G,2017ApJ...840L..24F,2018PhRvL.120o1101R}: 
they should be more massive than BHs born from 
``ordinary'' stellar evolution; they should merge later, at lower redshift, than earlier generation BHs;   and after a few mergers their dimensionless spin magnitudes (on average) should cluster  around  $\approx 0.7$ \footnote{A dimensionless spin magnitude of one corresponds to a maximally spinning BH, while zero spin magnitude corresponds to a non-spinning BH.}
while if the holes have gained a significant fraction of their mass via minor mergers
their spin should be low \citep[e.g.,][]{2002ApJ...581..438M}.

During the inspiral  of a binary BH, gravitational wave (GW) radiation is emitted anisotropically due to asymmetries in the merger configuration.   Numerical relativity simulations show that the kick velocity for  non-spinning  objects depends on the mass ratio of the binary 
$q=m_2/m_1\le 1$ as $v_{\rm GW}\approx 175\,{\rm km\,s}^{-1}f(q)$, where $f(q)=68.5\,\eta^2(1-4\eta)^{1/2}(1-0.93\eta)$ and $\eta\equiv q/(1+q)^2$,
which peaks at $f(q\approx0.36)=1$ and vanishes at $f(0)=f(1)=0$.
But, if the BHs have a finite spin, 
the kick could be much larger \citep{2006PhRvD..73l4006D,2007PhRvL..98w1101G,Campanelli2007}. 
Generally, 
whether significant BH growth  occurs or not will depend
on the host cluster physical properties.
{ \citet{2018PhRvL.120o1101R}  set the initial spin of the BHs to zero and showed that in this 
case  second-generation BHs  can form in clusters with properties similar to current day globular clusters. 
However, the second generation BHs are promptly ejected after a second merger, so that no  growth above about $100M_\odot$ can ever 
occur in globular clusters.
}

In this paper, we compute 
the  mass of the largest BH that can be formed through  hierarchical mergers of smaller BHs, and determine how its value is set by the global properties (i.e., mass and averaged density) of the host cluster, including a prescription for the dynamical evolution of the cluster itself.
The possibility of significant  
BH growth within a dense star cluster has been
explored before \citep[e.g.,][]{1987ApJ...321..199Q,2011ApJ...740L..42D,2014MNRAS.442.3616L}. 
Most of this previous work focused on the evolution
of extremely dense clusters where a relativistic instability sets in, driving the formation
of a massive seed via the ``catastrophic'' collapse of the cluster
core.
For this collapse to occur, however,
the merger timescale of binaries due to the emission of gravitational radiation must be  shorter  than the timescale for cluster heating via binary-single encounters.
This condition typically requires a cluster velocity dispersion 
$\gtrsim 1000\,\rm km\ s^{-1}$ \citep[e.g.,][]{1995MNRAS.272..605L,2006MNRAS.371L..45K}, which is much larger than that measured  in
present-day star clusters.

Here, we work under the more conservative and realistic assumption that 
the dynamical  hardening of BH binaries is an
efficient energy source which  complies with H\'{e}non's principle \citep{1975IAUS...69..133H}. 
According to this principle,  the rate of energy generation 
in the core is regulated by the energy demands of the bulk of the system. {\citet{2013MNRAS.432.2779B} showed that a population of stellar-mass BHs can act as the central energy source, implying that there is thermal contact between the BHs and the stars. As a consequence, a BH population can be dynamically retained for as long as 10 half-mass relaxation timescales of the  cluster. } For the systems we consider here, the bulk of the system is in the light stellar component, while the energy flux must be supplied by the BHs, and ultimately must be generated by hard binaries in the core of the BH subsystem \citep[e.g.,][]{2013MNRAS.432.2779B}.
 H\'{e}non's principle allows us  to relate the 
hardening rate of the binaries to the {\it global} properties of their parent cluster, 
as opposed to the {\it core} properties as done in the previous literature
\citep[e.g.,][]{2004ApJ...616..221G,2016ApJ...831..187A,2018MNRAS.481.4775D,2018arXiv180901164C}. The advantage is that 
the {energy generated by the binaries in the core can be easily linked to the secular evolution of the cluster}, and thus a complete description of the cluster and binary evolution can be obtained. 

In Section 2, we describe the analytical 
model that we used to compute the hardening rate
of BH binaries and the evolution  of the cluster in which they reside. We use this model to study the mass growth of a BH seed 
undergoing repeated mergers with smaller objects.
In order to treat this problem analytically we first work under the simplifying assumption that dynamical and GW recoil kicks can be ignored.
In Section 3, we add these and other additional ingredients to our models using
a semi-analytical method.
We adopt this method to follow the long term evolution of
BHs in dense star clusters and 
to derive the properties of the merging binaries they produce.
 The astrophysical implications
of our results are discussed in Section 4.

\section{Binary heating}\label{1rec}
Let us assume that the heating rate of BH binaries is so high that we need to invoke a ``capped" heating rate, which is set by the maximum heat flow that can be conducted outward by two-body relaxation. {This condition may not be met soon after the formation of the cluster, but because the dynamical friction time of BHs is short and BH binary formation is efficient, heating by BHs quickly goes up to generate the maximum heat flow.} This means that the heat production in the centre by BHs is  balanced by the global energy flow \citep{1961AnAp...24..369H,2011MNRAS.413.2509G,2013MNRAS.432.2779B}
 and we can relate the heat generation to the  cluster properties:
 
\begin{equation}
\dot{E} = \zeta\frac{|E|}{\trh}, 
\label{eq:Edot}
\end{equation}
where $E\simeq-0.2GM_{\rm cl}^2/\rh$ is the total energy of the cluster, with $M_{\rm cl}$ the total cluster mass and $\rh$ the half-mass radius. The constant $\zeta\simeq0.1$ \citep{2011MNRAS.413.2509G,2012MNRAS.422.3415A},
 and $\trh$ is the average relaxation time-scale within $\rh$ which is given by \citep[e.g.,][]{1971ApJ...164..399S}
 \begin{equation}
\trh = 0.138\sqrt{\frac{M_{\rm cl}\rh^3}{G}}\frac{1}{\mmean\psi\ln\Lambda}.
\end{equation}
Here $\mmean$ is the mean mass of the stars {and stellar remnants} and $\ln\Lambda$ is the Coulomb logarithm, which depends weakly on the total number of stars;  we ignore this dependence  and set $\ln\Lambda=10$. The constant $\psi$ depends on the mass spectrum within $\rh$, and is often assumed to be $\psi=1$, which applies to equal-mass clusters. For a full mass spectrum, however,  $\psi$ can be as high as $\psi\simeq30-100$ \citep{2010MNRAS.408L..16G}. 
{ In what follows, we adopt $\psi=5$. This  takes into account that in the early evolution of the cluster
 (first $\sim100$Myr) the mass function contains more massive stars and $\psi$ is high.
After this time, as a result of the BHs, the mass function remains wide despite the turn-off mass being $\lesssim 1 M_\odot$.
In the case of equipartition, Eq. (24) in \citet[][]{1971ApJ...164..399S} states: 
$\psi = 
\langle m_\star^{2.5}\rangle/\langle m_\star\rangle^{2.5}$.
If we approximate the cluster by two mass components, BHs with mass 
$10 M_\odot$ and stars with mass $0.6 M_\odot$, then for a mass fraction in 
BHs of a few percent  we find  $\psi \simeq 5$.}

Combining the two expressions above we find for the heating rate for $\mmean=0.6\ M_\odot$,
$\zeta=0.1$ and $\psi=5$
\begin{align} \dot{E}_0 \simeq\, &2.3\times10^5\,\msun(\kms)^2\,\myr^{-1}M_5^{2/3}\rho_{5,0}^{5/6},
\label{eq:edot}
\end{align}
where the subscript 0 refers to initial values and 
we expressed the result in terms of $M_5 = M_{\rm cl}/10^5\,\msun$ and $\rho_{5,0} = \rho_{\rm h,0}/10^5\,\msun/\pc^3$, with $ \rho_{\rm h,0}= 3M_{\rm cl}/(8\pi r_{\rm h,0}^3)$ the averaged density within the cluster half-mass radius. 

In the absence of other effects, the cluster has a constant radius until $t=t_0$, where $t_0$ is the time for BHs to reach the centre and start heating the cluster.
{This is} a fraction of the initial $\trh$, { {i.e., $\lesssim 1$\,Myr} for the model parameters used about}, {hence} we {can safely} assume that heating starts immediately and therefore $t_0=0$. {In Section~\ref{3rec} we include larger $t_0$, because for the most massive clusters $t_0$ can be $\gtrsim100\,$Myr.}
If we neglect mass loss from stellar evolution and the escape of BHs {and stars} {(i.e., we assume a constant $M_{\rm cl}$)}, then the evolution of the cluster radius follows from the energy evolution and the assumption of virial equilibrium: $\rhdot/\rh = \dot{E}/|E|$ and solving this gives \citep{Henon65}

\begin{align}\label{rex}
\rh(t) = 
\displaystyle\rhn \left(\frac{3}{2}\frac{\zeta t}{\trhn}+1\right)^{2/3}, 
\end{align}
such that 
\begin{align}\label{vex}
\vesc(t) = 
\displaystyle\vescn \left(\frac{3}{2}\frac{\zeta t}{\trhn}+1\right)^{-1/3}, 
\end{align}
where 
\begin{align} 
\vescn &\simeq 50\,\kms\, M_{5}^{1/3}\rho_{5,0}^{1/6}, \label{vescF} 
\\
\trhn &\simeq 7.5\,\myr\, M_{5}\rho_{5,0}^{-1/2}.
\end{align}
The constant of proportionality in the escape velocity applies to  a King (1966) model with $W_0=7$. 
Substituting these relations in Eq.~(\ref{eq:edot}) we find
\begin{align}
\dot{E}(t) \simeq\, &\dot{E}_0 \left(\frac{3}{2}\frac{\zeta  t}{\trhn}+1\right)^{-5/3}\ .
\label{eq:edott}
\end{align}
The rapid decrease of $\dot{E}$ with time is the result of cluster expansion, increasing $\trh$ and lowering $E$ {(see Eq.~\ref{eq:Edot})}.

Assuming that the heating is produced 
by the BH binaries in the core, it follows that  the  rate at which 
the core binaries harden is\footnote{{We note that in balanced evolution $|\dot{E}_{\rm bin}|$ is actually slightly larger than $\dot{E}$, because some of the binary energy is used to eject stars/BHs with velocities in excess of the local escape velocity. The removal of the binding energy of the star/BH contributes indirectly to the heating of the core, but the excess energy above the escape energy does not contribute to the heating. For single-mass clusters, $|\dot{E}_{\rm bin}|\simeq1.1\dot{E}$ \citep{Goodman84}. 
}}
\begin{align}
\dot{E}_{\rm bin} =-{\dot{E}} \ ,
\label{eq:edotbin2}
\end{align}
where $E_{\rm bin}=-Gm_1m_2/2a$,  and $m_1$ and
$m_2$  are  the masses of the 
binary components. 
Here
we have assumed that at any time one binary is responsible
for most of the heat production in the cluster core.

We can compare our results to the scaling of the hardening rate often used
in previous work \citep[e.g.][]{Heggie2003,Binney2011}: $\dot{E}_{\rm bin} = 0.2 E_{\rm bin}/\tau_{\rm enc}$, where $\tau_{\rm enc}^{-1}\simeq8\pi G \rho a/\sigma$ is
the binary-single encounter timescale, 
and $\rho$ and $\sigma$ are the density 
and velocity dispersion of the BHs near the binary, respectively.
{ 
A choice often made in the literature 
is to  set $\rho$ equal to the density of stars
in the core
\citep{2004ApJ...616..221G,2006ApJ...640..156G,2016ApJ...831..187A,2018MNRAS.481.5445S,2017arXiv171107452S,2018arXiv180901164C}. 
This choice, however, is quite arbitrary as the core density of the BHs
is not known a priori and  cannot
be easily linked to the cluster global properties.
For the King model used above, taking $m_1=m_2=10M_\odot$ 
 one finds  $\dot{E}_{\rm bin} \simeq -\pi G^2m_1m_2\rho/\sigma = -7\times 10^2\msun(\kms)^2\,\myr^{-1} M_{\rm cl}^{-1/3}\rhoh^{5/6}$. }
Our assumption of balanced equilibrium
 instead led to the different normalisation and scaling
$\dot{E}_{\rm bin}\simeq -2.3\times10^5\msun(\kms)^2\,\myr^{-1} M_{\rm cl}^{2/3}\rhoh^{5/6} $.

The much lower value for the binary heating rate used in some previous studies implies 
a non-equilibrium state where the rate of energy generation by the core binaries 
 cannot sustain the cluster core against  collapse.
In this situation,
the BH core will rapidly contract causing
the  binary-single interaction rate 
 to increase until $\dot{E}_{\rm bin}$ 
matches the value
given by our
Eq.\ (\ref{eq:edotbin2}), after which
the balanced evolution will be restored.
This fits in the view  that the rate of flow of energy is controlled by the system as a whole, not by its core 
properties.

Eq.~(\ref{eq:edotbin2}) can be equivalently expressed in
terms of the binary semi-major axis,
\begin{align}
\dot{a}_{\rm } ={2a^2\over Gm_1m_2} {\dot{E}_{\rm bin}}\ .
\label{eq:edotbin}
\end{align}
The  lifetime of a BH binary is  then
\begin{align}
\tau_{\rm bin}=
\int_{a_{\rm m}}^{a_{\rm h}}
{Gm_1m_2\over2a^2}\dot{E}_{\rm bin}^{-1}da\simeq 
-{Gm_1m_2\over2a_{\rm m}}\dot{E}_{\rm bin}^{-1},
\label{eq:thard}
\end{align}
with $a_{\rm h}\approx G\mu/\sigma^2$ the semi-major axis at the hard/soft boundary, 
where $\mu=m_1m_2/(m_1+m_2)$,
and 
$a_{\rm m}$ {the semi-major axis}
 at which the sequence of hardening interactions terminates because
either the binary merges or it is ejected from the cluster. 
This latter quantity is determined by
\begin{align}\label{am}
a_{\rm m}=  \max \left(a_{\rm GW},a_{\rm ej}\right) \ ,
\end{align}
with $a_{\rm GW}$ and $a_{\rm ej}$ defined below.
{ Note that when solving Eq.\ (\ref{eq:thard}) 
we have taken $\dot{E}_{\rm bin}$  out of the time integral, i.e.,
we have assumed that  the cluster properties do not change over $\tau_{\rm bin}$.
This is justified because $\vert E/E_{\rm bin}\vert\sim M_{\rm cl}/(m_{1}+m_2)\gg1$.}
Moreover, we have used the fact that $a_{\rm m} \ll a_{\rm h}$.

{As the binary  binding energy 
increases following  a binary-single interaction, from energy 
and momentum conservation one finds that the centre of mass receives a recoil kick
with velocity \citep{2009ApJ...692..917M} 
\begin{align}\label{vkn}
v_{\rm 2-1}^2\simeq 0.2 G {m_1m_2\over
  {m_{123}}} {q_3\over a} 
 \ ,  \end{align}
where $q_3=m_3/(m_1+m_2)$, with $m_3$ the mass 
of the interloper, and $m_{123}=m_1+m_2+m_3$.}
In deriving Eq.\ (\ref{vkn})
we have assumed that during each
binary-single interaction the
binding energy  of the binary 
increases  by a fixed fraction
$\delta\approx 0.2$ \citep[e.g.,][]{1996NewA....1...35Q,2002MNRAS.330..232C}.
It follows that the binary is ejected from the cluster when its semi-major  axis  drops below \citep{2016ApJ...831..187A}:
 \begin{eqnarray}\label{aej}
a_{\rm ej}=  0.2 G {m_1m_2\over m_{123}} {{{q_3}}\over{v_{\rm esc}^{2}}}
\simeq0.1 {\rm AU}\left({m_1m_2\over m_{123}} q_3 {0.6 \over M_\odot}  \right) \left(50 {\rm km\ s^{-1}}\over v_{\rm esc}\right)^{2} .
\end{eqnarray}

In reality, for the high velocity dispersion clusters
considered here, most (if not all)
binaries will merge before $a$ has decreased 
to $a_{\rm ej}$.
Then, the semi-major axis at which a merger
will occur is determined by requiring that the
rate of energy loss due to dynamical hardening 
equals that due to GW radiation
\citep{Peters1964}:
\begin{align}\label{2.5}
\dot{a}|_{\rm GW}=
-{64\over 5}\frac{G^3m_1m_2(m_1+m_2)}{c^5a^3(1-e^2)^{7/2}} g(e),
\end{align}
with $c$ the speed of light, $e$ the binary eccentricity, $g(e)=\left(1+{73\over24}e^2+\frac{37}{96}e^4\right)$,
and the merger time is 
$\tau_{\rm GW}=a/\vert\dot{a}|_{\rm GW}\vert$. Solving 
 $\dot{a}|_{\rm GW}=\dot{a}$,  gives  
 \begin{align}\label{agw}
 a_{\rm GW}=  \left(-{32\over 5} 
{G^4 (m_1m_2)^2(m_1+m_2)g(e)\over c^5 (1-e^2)^{7/2} }   
\dot{E}_{\rm bin}^{-1}\right)^{1/5} \nonumber\\
\simeq10^{-2}\ {\rm AU} \left(
{{(m_1m_2)^2(m_1+m_2)\over M_\odot^5} }{M_\odot(\kms)^2\,\myr^{-1}\over 
\dot{E}_{\rm bin}}
\right)^{1/5} \nonumber \\
\times\left({g(e)\over (1-e^2)^{7/2}}\right)^{1/5}
; 
\end{align}
{ at  $a_{\rm GW}$ the rate of orbital evolution due to binary-single encounters 
is comparable to that due to GW emission.} 
{ Note that that for clusters
in which dynamical ejections can be ignored,
 Eq.\ (\ref{agw}) and Eq.\ (\ref{eq:thard}) show that
the binary hardening rate and lifetime are independent of $m_3$.
}

{In the next section we will use the model described above to address under which conditions the merger of stellar-mass BHs in a dense star cluster can lead to the rapid growth of a massive seed.
}

\begin{figure}
\centering
 \includegraphics[width=3.3in,angle=0.]{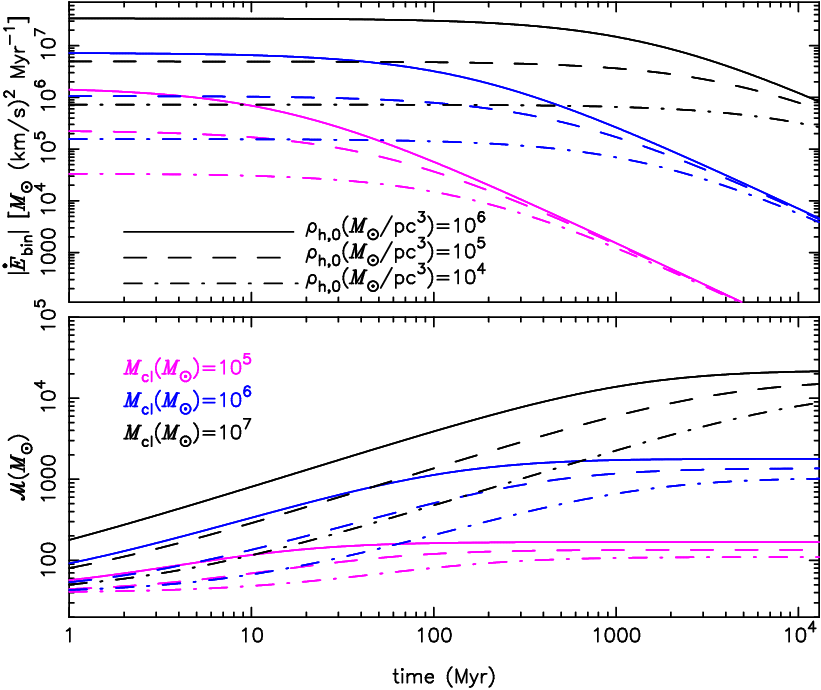}
 \caption{
Time evolution  of 
 $\dot{E}$ and
$\mathcal{M}$ as a function 
of cluster mass and density.
In the bottom panel we
set $m_2=\mathcal{M}_0=20\msun$ and
$e=2/3$ 
\citep[the mean value of 
a thermal distribution;][]{1919MNRAS..79..408J}.
  }
 \label{fig1}
 \end{figure}

\subsection{Mass growth of
a central seed
}\label{12rec}

{ There are (at least) two mechanisms that could lead to the ejection of {BHs} 
from a cluster:
(i) binary-single interactions in which either one BH, or both the binary and the single BH escape {as the result of momentum conservation};
and (ii) the {relativistic} momentum kick of the merger due to the anisotropic emission of GW
radiation.  
We consider here an analytical model where these two mechanisms
are ignored. 
This more tractable and idealized model will provide some important insights on the behavior
of real clusters,  that are independent of the assumptions made.
Moreover, it is a particularly good approximation  in at least two cases of interest: for clusters with large escape
velocities, and for $q\ll1$ leading to low recoil kicks.}

If we assume that the ejection mechanism (i) can be ignored,
then  the overall merger rate of BH binaries
is simply given  by
\begin{align}\label{rate}
\Gamma_{\rm m}\simeq 
{\tau_{\rm bin}^{-1}}={2a_{\rm GW} \over Gm_1m_2}{\vert\dot{E}_{\rm bin}\vert} \ .
\end{align}
Hence,  the merger rate of BH binaries produced dynamically inside a massive cluster
is approximately proportional to 
the hardening rate of the core binaries and  scales
with the cluster properties as $\Gamma_{\rm m}\propto M_{\rm cl}^{8/15}\rho_{\rm h}^{2/3}$. 
{ Note  that the latter expression gives the BH merger 
rate even in the presence of relativistic kicks, but only until the escape velocity of the cluster is sufficiently 
large that mechanism (i) can be ignored,
which requires $v_{\rm esc} \gtrsim 100\,\rm km\ s^{-1}$
\citep{2016ApJ...831..187A}.
This condition is often met
in present-day nuclear star clusters, 
and might also have been present
in smaller clusters if they were initially much denser than observed today 
(see also Section\ \ref{astro}). }

A natural question to ask is whether significant 
growth can ever occur on
a sufficiently short timescale. 
If we now
neglect both ejection
mechanisms (i) and (ii) above,
then
Eq.\ (\ref{rate}) 
leads to the following upper limit to the mass growth rate associated with a BH undergoing repeated mergers with 
  smaller BHs of mass $m_2$: 
 \begin{align}\label{Mrate}
 \dot{\mathcal{M}} =\Gamma_{\rm m}m_2 \ .
\end{align}
Clearly, this latter expression should be interpreted as 
an upper limit to the real growth 
rate because we are neglecting dynamical and GW kicks. Such kicks 
will slow down the growth process by removing temporarily the hole from the cluster core, or halt it by fully removing the BH from the cluster when the
recoil is sufficiently large.
These additional complications will be considered in the analysis of the following section.

Integrating  Eq.\ (\ref{Mrate}), 
we find the BH seed growth equation
 \begin{align}\label{Mrate2}
{\mathcal{M}}(t) \simeq&
\mathcal{M}_0+
\Big[ {42\over 5G} {K_{\rm GW}^{1/5}m_2^{2/5}
\vert\dot{E}_{\rm bin,0}\vert^{4/5}
   \nonumber
}\\ &
\times 
{2\trhn \over 3\zeta}
\left(1-\frac{1}{(3\zeta t/2\trhn
+1)^{3/2}}\right)
\Big]^{5/7} \ . 
\end{align}
where 
$K_{\rm GW}={32\over 5} 
{G^4 g(e) c^{-5} (1-e^2)^{-7/2} }$, and
$\mathcal{M}_0$ is the initial BH mass.
The
 growth timescale is
 \begin{align}\label{tgrow}
 \tau_{\rm M}=
{2\trhn \over 3\zeta}\left[\left(1-{5   G {\left(\mathcal{M}-\mathcal{M}_0\right)}^{7/5} \over  42  K_{\rm GW}^{1/5} m_2^{2/5}\vert\dot{E}_{\rm bin,0}\vert^{4/5}} {3\zeta\over 2 \trhn}\right)^{-2/3}-1\right]\ .  ~~~~~~~
\end{align}

The mass growth predicted by
Eq.\ (\ref{Mrate2}) is shown in
Figure\ \ref{fig1} as a function of cluster mass and 
central density (bottom panel); the upper panel of the figure gives
the evolution of the 
binary hardening rate. 
From Figure\ \ref{fig1} we see that
the growth of the seed mass
is interrupted 
 soon after the
relaxation driven expansion of the
cluster starts, causing 
 $\vert\dot{E}_{\rm bin}\vert$
to decrease. 
The physical reason that the merger rate drops, is because the energy demand decreases, and significant drop in $\dot{a}$ happens after one initial relaxation time, which is $\propto M_{\rm cl}/\sqrt{\rhoh}$.
In fact, Eq.\ (\ref{Mrate2}) implies
that for
$t\rightarrow \infty$, ${\mathcal{M}}(t)$ takes the finite value
\begin{align}\label{Mmax3}
{\mathcal{M}}_{\rm max}
=\mathcal{M}_0+110\ \msun M_5^{23/21}\rho_{5,0}^{5/42}
m_{20}^{2/7},
\end{align}
with $m_{20}=m_{2}/20\msun$.
Eq. (\ref{Mmax3})
demonstrates the important result that
the maximum mass that can be attained   via hierarchical 
mergers is set by the mass of the
host cluster, and it is approximately independent
of the cluster density, the seed mass $\mathcal{M}_0$ and the mass of the accreted BHs. 

{
The model presented here 
 leads to the following  conclusions.
First, Eq. (\ref{Mmax3}) shows that 
 a minimum mass
$M_{\rm cl}\gtrsim 10^6M_\odot$ is
required for the formation of a
 BH with mass $\mathcal{M}\gtrsim 10^3M_{\odot}$.
This
implies that massive clusters 
are the only places in which significant growth can
ever occur through repeated
BH collisions.
Second, the lack of dependence  on $\mathcal{M}_0$ and $m_2$ shows that the growth 
 process should be largely independent of
 the  the initial BH mass distribution and therefore of metallicity. 
 These conclusions are 
supported by the more
detailed  models described 
in the following section, which include additional 
physics that we have so far ignored, e.g., dynamical friction,
natal kicks,  GW and dynamical recoil kicks. {Including these effects increases the importance of the initial cluster density.}
The main results of our work are derived from these  more detailed models.
}

\section{Method}\label{3rec}
We simulate the evolution of the cluster and of its BH
population using 
a semi-analytical technique.
Briefly,  we assume that at any time
a BH binary exists in the
cluster core 
and that this binary hardens
at a rate given by Eq.~(\ref{eq:edotbin}).
Thus, each binary remains in the cluster core 
for a time $\tau_{\rm bin}$, and 
after this time a new binary is formed.
We evolve the cluster BH population
until either all BHs have been ejected, or
until a time of $13\,\rm Gyr$ has passed.
Additional details of our method are
given  in the following.

As a first step  we
 initialise the cluster model.
This is determined
by two parameters: the initial 
density,
$\rho_{\rm h,0}$, 
and the cluster mass, $M_{\rm cl}$.
All time dependent quantities are derived from the analytical model
of Section~\ref{1rec}.

Then, we set the initial 
mass and spin distributions of the BHs. 
The initial spins of the BHs
are drawn  uniformly from
the range $\chi=cJ/(Gm^2) = 
\left[0, 1\right)$, with $J$
the spin angular momentum and
$m$ the BH mass. In order to address how our results depend on the
assumed spin distribution, we  also explore other choices
for the BH spins, setting their initial values all
 to $0$  or $0.8$.
  { 
When a BH remnant  is retained and merges  with new BH companions, we then track
the evolution of its  spin magnitude forward in time using the prescriptions in 
\citet{2008PhRvD..78d4002R}}.
 We sample the masses
of the
stellar  progenitors from a Kroupa initial mass function  
\citep{Kroupa2001} with  masses in the range
$20$ to $100M_\odot$, and evolve the stars  to BHs 
using the Single Stellar Evolution (SSE) package  \citep{Hurley2002}. 
{ We consider the lower limit of $20M_\odot$ because
only stars with masses larger than this produce BHs.}
Unless otherwise specified,
we take a metallicity $Z=0.1Z_\odot$. But note
that for the stellar mass range considered,
metallicities lower than this would all produce 
similar initial BH mass functions  \citep[e.g.,][]{2017MNRAS.470.4739S}.
Our models adopt the updated prescriptions for stellar
winds and mass loss, in order to replicate the BH
mass distribution of 
\citet{Dominik2013} and \citet{Belczynski2010},
but include updated prescriptions for the pulsational pair-instability in massive stars \citep{Belczynski2016}. 
In our simulations, no BH can be born with a mass above $40M_\odot$.
Our initial sample contains a number
of BHs equal to $0.2\%$ of the total 
number of stars in the cluster \citep[typical for a standard  initial mass function; e.g.,][]{2010MNRAS.402..519L}.
For this choice we find 
that about $4\%$ ($2\%$) of the total cluster
mass is in BHs for $Z=0.1Z_\odot$ 
($Z=Z_\odot$).

For each BH we compute a natal kick velocity from a Maxwellian
with dispersion $265 \kms$, as commonly done for neutron
stars \citep{Hobbs2005}, { and assume that 
 the momentum imparted on a black hole is the same as the momentum given to a 
 neutron star \citep{Fryer2001}. Thus,
the natal velocity of a
BH is lowered by the factor of $1.4M_\odot /m$, with $1.4M_\odot$ the typical 
neutron
star mass.}
 From this initial sample we remove those BHs which received a natal kick
with velocity larger than $v_{\rm esc,0}$. 
{ Using this prescription, and for
the high velocity dispersion clusters we consider below ($v_{\rm esc,0}\gtrsim 70\rm km\ s^{-1}$),
the fraction of  ejected BHs to their initial number is in all cases
$\lesssim 0.01$.  Note that the fraction of ejected
BHs  depends on the {\it initial} escape velocity which is related to the 
cluster mass and density  through  Eq.\ (\ref{vescF})
}.

\begin{figure*}
\centering
 \includegraphics[width=3.4in,angle=0.]{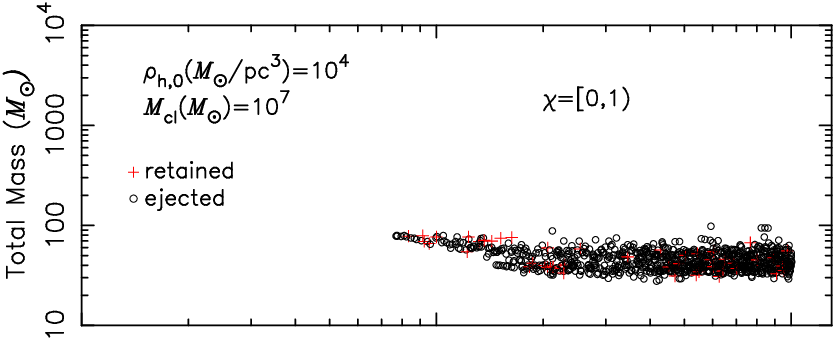}
 \includegraphics[width=3.3in,angle=0.]{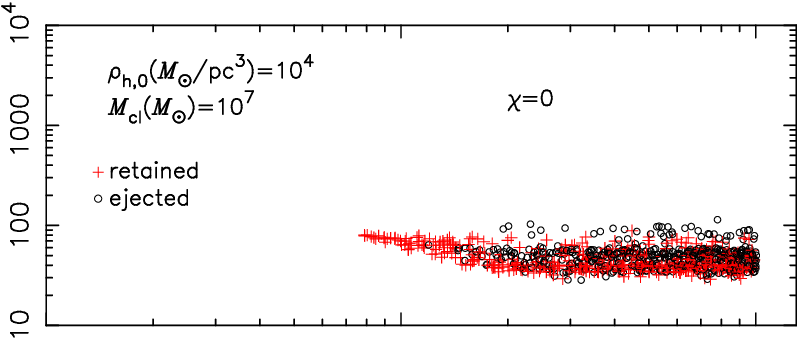}
 \includegraphics[width=3.4in,angle=0.]{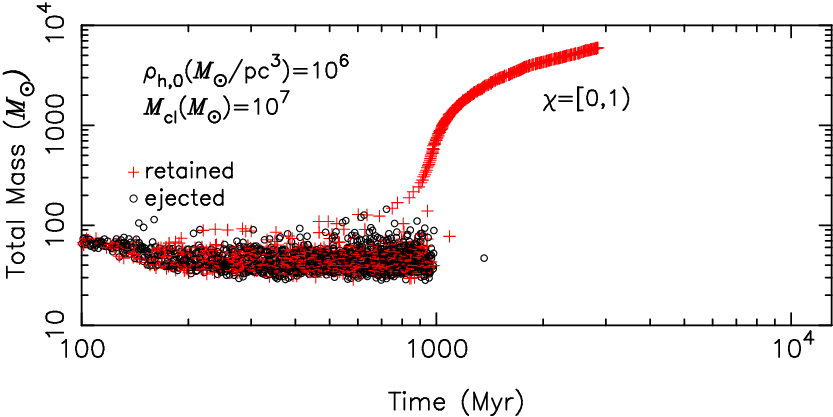}
 \includegraphics[width=3.3in,angle=0.]{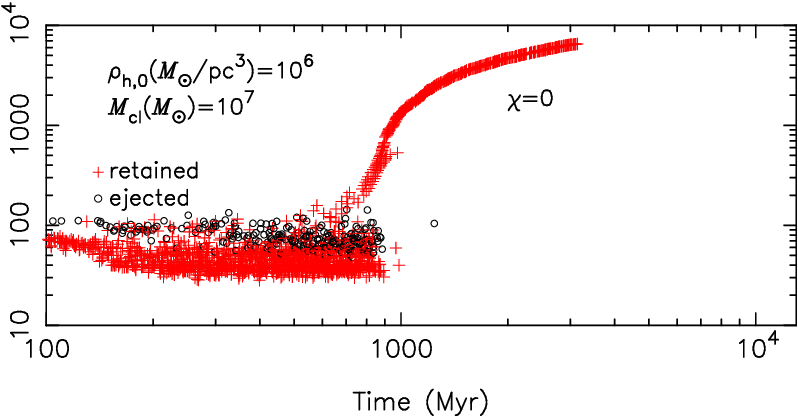}
 \caption{Mass of the merging BH binaries formed in four example
 clusters. We consider two  models with  the same total mass but different densities to illustrate
 the effect of the latter on the evolution of the BH population.
 In the left panels the BH spins  are  distributed uniformly between 0 and 1; in the right panels the BHs have no spin initially.
 Red circles are BH remnants which are retained inside
 the cluster, following a relativistic kick. 
  }
 \label{fig3}
 \end{figure*}
 
 \begin{figure*}
\centering
\includegraphics[width=3.5in,angle=0.]{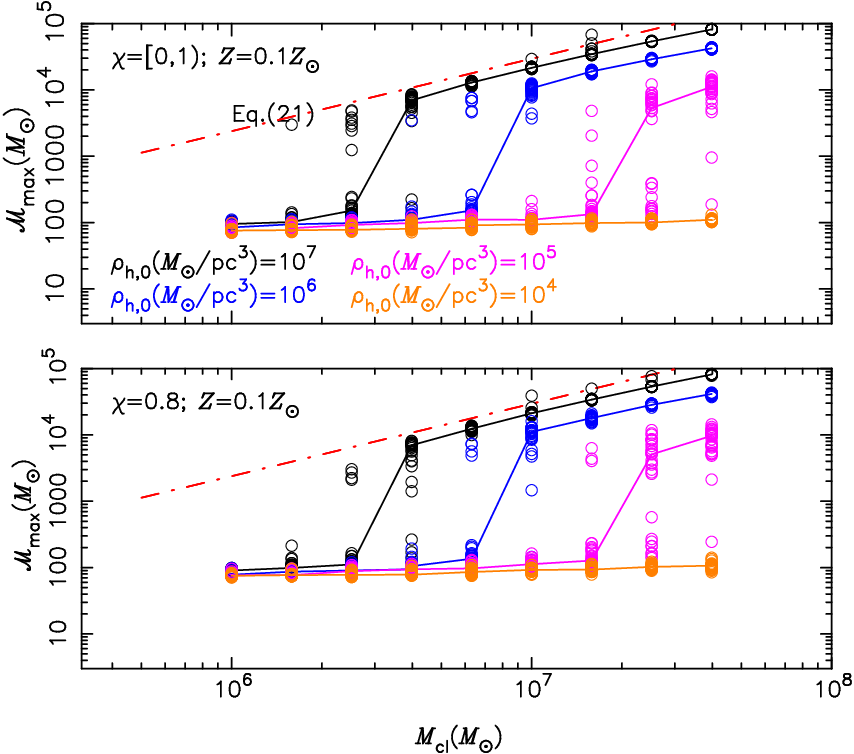}
\includegraphics[width=3.35in,angle=0.]{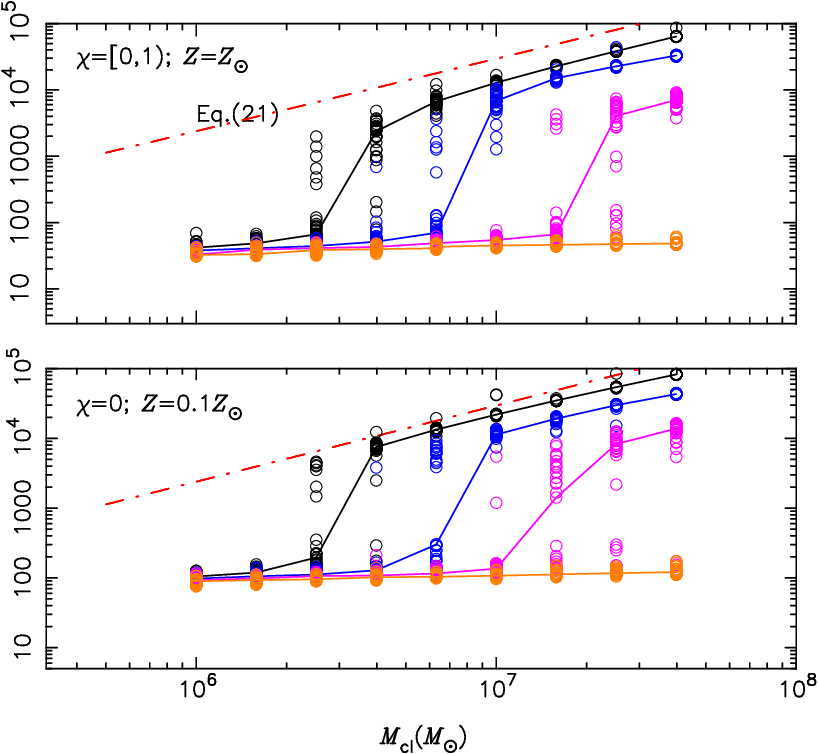}
 \caption{Maximum mass, $\mathcal{M}_{\rm max}$, of the merging BH binaries formed over 13Gyr of evolution,
 as a function of cluster properties. For each value of cluster density and mass we display the results from 30 random cluster realisations.
 Solid lines are median values.
In the top-right we set $Z=Z_{\odot}$, and in the lower panels we set the initial BH spin magnitudes
 to $0.8$ (left) and $0$ (right).
 A comparison of these latter models
 to our fiducial set with $Z=0.1\times Z_{\odot}$
and initial spins in the range $\chi = 
\left[0, 1\right)$ (upper left) shows that
assumptions on spins and
metallicity have little effect on the value of the cluster mass
and density at which substantial BH growth occurs.
  }
 \label{fig-maxmass}
 \end{figure*}

  \begin{figure*}
\centering
\includegraphics[width=3.5in,angle=0.]{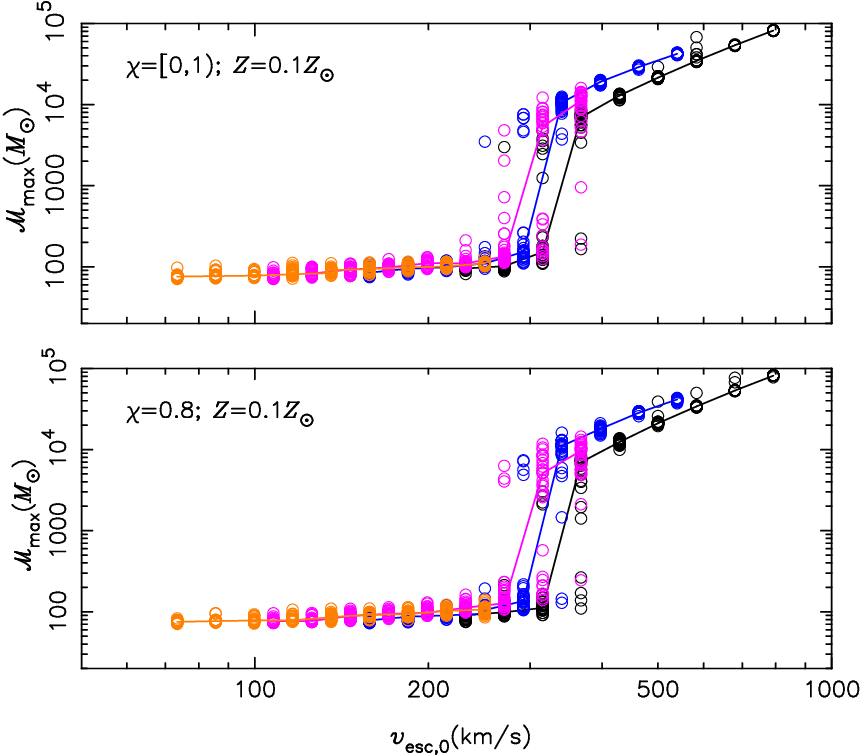}
\includegraphics[width=3.35in,angle=0.]{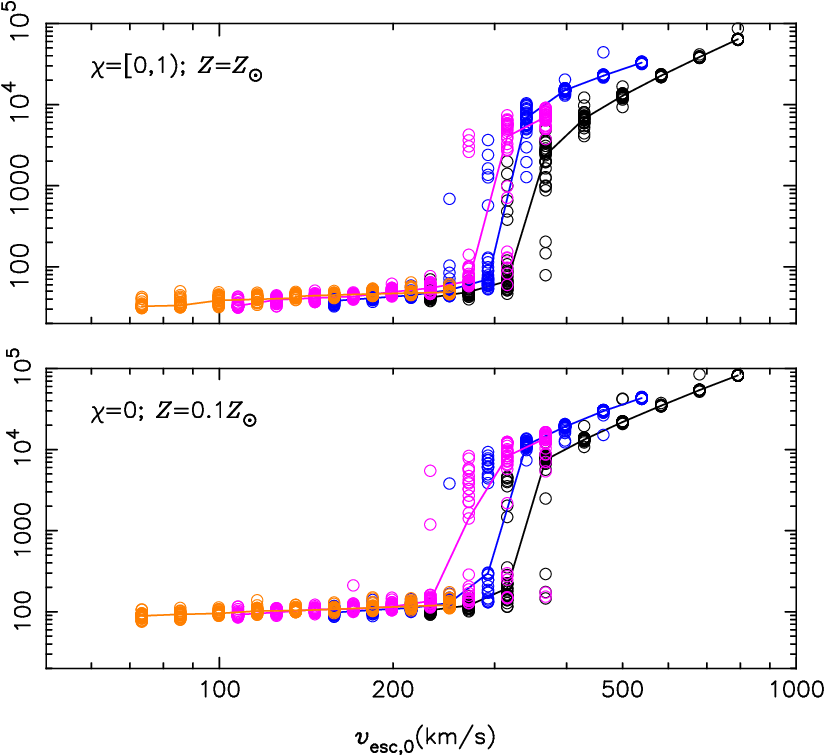}
 \caption{Similar to Fig. \ref{fig-maxmass},
 but  now the  mass of the largest BH formed over 13 Gyr, $\mathcal{M}_{\rm max}$, is plotted
 as a function of cluster initial escape velocity.
 Different colors correspond to different densities as
 indicated in Fig. \ref{fig-maxmass}.
  }
 \label{fig-vesc}
 \end{figure*}

 \begin{figure}
\centering
\includegraphics[width=2.6in,angle=0.]{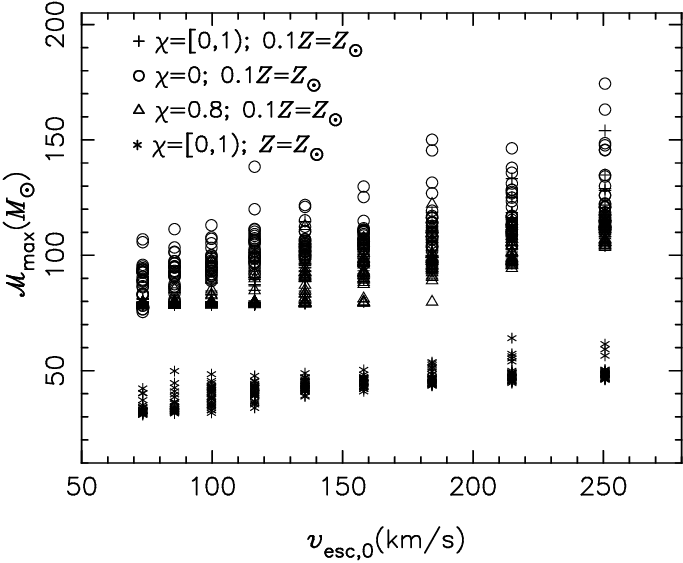}
 \caption{Zoom-in  of the results displayed in Fig.\ \ref{fig-maxmass},
 showing the largest BHs formed in clusters with initial
 escape velocity $\lesssim 300\rm km\ s^{-1}$
 and $\rho_{\rm h,0}=10^4\ M_\odot\rm\ pc^{-3}$. In these  clusters, any growing BH is ejected
 after a few mergers, implying  that the maximum mass that can be built though hierarchical mergers
is affected significantly by the initial mass function of the BHs
 and their spins. 
 }
 \label{zoom}
 \end{figure}

At the start,
we assign to each BH  a dynamical friction time
scale \citep{Binney2011}:
\begin{equation}\label{dfr}
\tau_{\rm df} \simeq 1.65\,r_{\rm h}^2{\sigma\over {\ln \Lambda G m}}=4.5\,{\rm Myr}\ \left(m\over {20\,M\odot}\right)^{-1}
M_{5}\rho_{5,0}^{-1/2}
\end{equation}
where $\sigma$ is the  cluster one-dimensional velocity dispersion,
and $v_{\rm esc}\simeq4.77 \sigma$ for the adopted cluster King model.
After this time has passed, a BH enters
the core region and is allowed to form a binary
with another BH. 
After a time $t_{\rm bin}$ a new binary is formed
 assuming that the pairing probability for its BH components scales as $\propto 
(m_1+m_2)^4$,  appropriate for binaries formed via three body processes
\citep{2016ApJ...824L..12O}.
Setting the  formation  time for the next binary equal to the hardening timescale  assures that at any time
a binary exists in the centre, providing the energy needed to sustain the cluster core against collapse.
 The  time from binary formation to  merger 
 is  set equal to
$\tau_{\rm merge}(a_{\rm m})=\tau_{\rm bin}+\tau_{\rm GW}$, where 
$a_{\rm m}$ is
obtained by sampling $e$ from a thermal
distribution,
$N(<e)\propto e^2$.

After  a merger, the  remnant BH receives a recoil kick as the result of the anisotropic emission of GW radiation with
velocity $v_{\rm GW}$ 
(typically, $v_{\rm GW}\gg v_{\rm 2-1}$). 
The subsequent evolution of the merger remnant depends on
whether it is retained or not in the cluster.
We compute $v_{\rm GW}$
using the fitting formula based on the results from numerical relativity simulations of
\citet{2008PhRvD..77d4028L},
\begin{equation}
{\vec{v}_{\rm GW}} = v_{\rm m}  {{\hat e}_{\perp,1}}+ v_{\perp} ({\rm cos} \,\xi \, {{\hat e}_{\perp,1}} + {\rm sin} \,\xi \, {{\hat e}_{\perp,2}}) + v_{\parallel} {{\hat e}_{\parallel}},
\label{eqn:kick}
\end{equation}
\begin{equation}
v_{\rm m} = A \eta^2 \sqrt{1 - 4\eta} \,(1 + B \eta), 
\end{equation}
\begin{equation}
v_{\perp} = {H \eta^2 \over (1 + q )} (\chi_{2\parallel} - q \chi_{1\parallel}),
\end{equation}
\begin{align}
v_{\parallel} = {16 \eta^2 \over (1+ q)} \left [ V_{1,1} + V_{\rm A} \tilde S_{\parallel} + V_{\rm B} \tilde S_{\parallel}^2 + V_{\rm C}\tilde S_{\parallel}^3 \right ] \times \nonumber
\\ 
|\, {\vec{\chi}_{2\perp}} - q {\vec{\chi}_{1\perp}} | \, {\rm cos}(\phi_{\Delta} - \phi_1),
\end{align}
where 
$\eta \equiv q/(1+q)^2$,
and $q=m_2/m_1$; $\perp$ and
$\parallel$ refer to vector components perpendicular and parallel to
the orbital angular momentum, respectively, and ${{\hat
   e}_{\perp,1}}$ and ${{\hat e}_{\perp,2}}$ are orthogonal
unit vectors in the orbital plane. 
The vector ${\vec{\tilde S}} \equiv 2({\vec{\chi}_2} + q^2 {
  \vec{\chi}_1})/(1+q)^2$.
The values of $A = 1.2\times 10^4\,\kms$, $B = -0.93$, $H = 6.9\times10^3\,\kms$, and $\xi = 145^{\circ}$
are  from \citet{2007PhRvL..98i1101G} and \citet{2008PhRvD..77d4028L}, and $V_{1,1} =
3677.76\,\kms$, $V_{\rm A} = 2481.21\,\kms$, $V_{\rm B} = 1792.45\,\kms$,
and $V_{\rm C} = 1506.52\,\kms$ are taken from  \citet{2012PhRvD..85h4015L}.
The angle $\phi_{\Delta}$ is that  between the
in-plane component ${\vec{\Delta}_{\perp}}$ of the vector ${
  \vec{\Delta}} \equiv (m_1+m_2)^2({\vec{\chi_2}} - q {\vec{\chi_1}})/(1+q)$ and the
infall direction at merger.  We take the phase angle $\phi_1$ of the binary to be random, and assume that
the spin directions during mergers are 
 isotropically distributed on the sphere.
We then assign a kick velocity $v_{\rm GW}$ to the merger remnant.

  \begin{figure}
\centering
\includegraphics[width=1.69in,angle=0.]{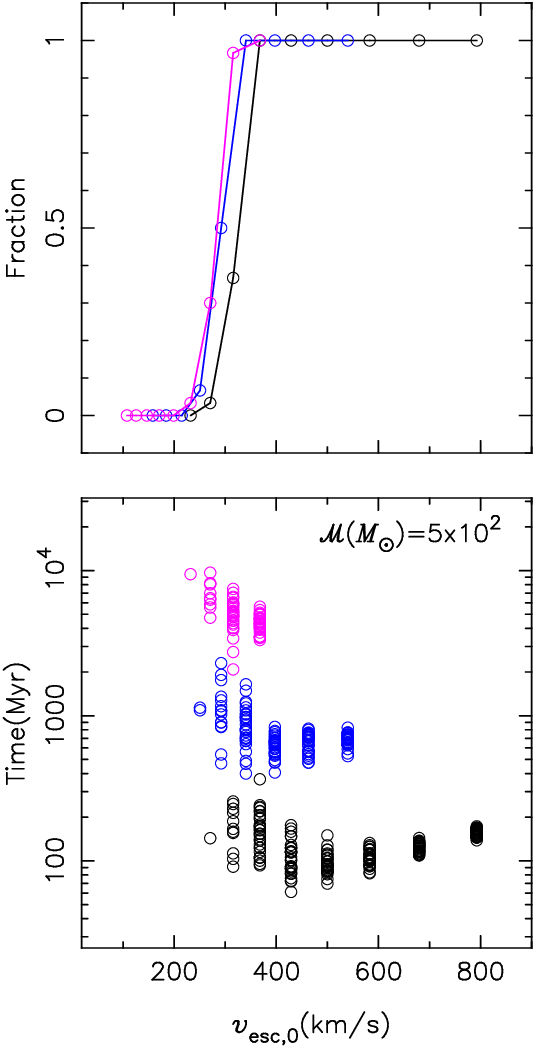}
\includegraphics[width=1.58in,angle=0.]{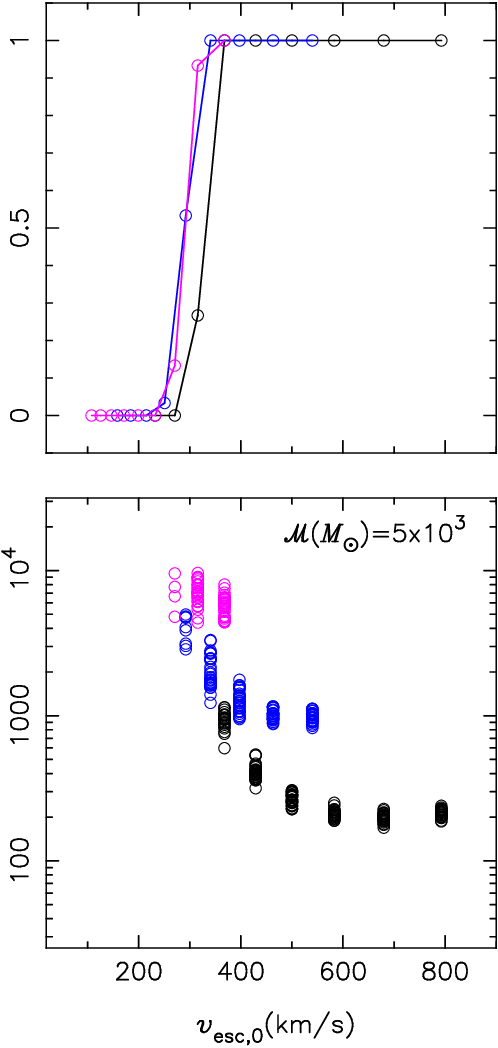}
 \caption{
Upper panels show the fraction of cluster models that are able to grow a BH of mass $5\times10^2$ (left) and $5\times10^3$ (right) over a total of 13\,Gyr of evolution.
In the lower panels we show the timescale at which these BHs are formed.
 For each value of cluster density and mass we display the results from 30 random cluster realisations.
 The figures refer to our models
 with $Z=0.1Z_\odot$ and initial spins distributed
between  $0$ and $1$.
  }
 \label{fig-fraction}
 \end{figure}
 
 The subsequent evolution has the following two possibilities:
 (i)  $v_{\rm GW}> v_{\rm esc}$.  
 In this case the remnant is
ejected from the cluster,  the remnant BH  is removed
from the sample, and a new binary is formed.
(ii)  $v_{\rm GW}< v_{\rm esc}$.
In this case, the remnant remains in the cluster,
we sum the masses of the two BHs and estimate the spin of
the remnant following the prescriptions in
\citet{2008PhRvD..78d4002R}. 
 The remnant is  deposited at a distance{\footnote{This was derived for a \citet{1911MNRAS..71..460P} model,
and it provides a good approximation also for moderately  concentrated King models  with $W_0 \lesssim 6$.
If the clusters are more centrally concentrated, then the excursions are smaller (by 1 order of magnitude for $W_0=8$) and 
$\tau_{\rm df}$ shorter. 
}}
$r_{\rm in}\simeq r_{\rm h} \sqrt{v_{\rm esc}^4/
(v_{\rm esc}^2-v_{\rm GW}^2)^2-1}$, sinks back to the core on the timescale 
$\tau_{\rm df}(r_{\rm in})$ and
 remains ``inactive'' until 
this time has passed. 
{ After a recoil kick,  the BHs will oscillate with decreasing amplitude, losing energy via dynamical friction;  $\tau_{\rm df}(r_{\rm in})$ provides an approximation of the time
over which the amplitude of the motion will fall to roughly the core radius.
} For a  detailed treatment
of this problem see \citet{2018MNRAS.474.3835W} (and references therein).
After a time  $\tau_{\rm df}$,
the remnant is
reintroduced in the core sample and is allowed to form a new binary. While the remnant stays inactive, new hard binaries are allowed to form in the core.

We terminate the integration if a time $t=13$\,Gyr
is reached, or all BHs have been
removed from the cluster,
{ including both the ejected binaries and the ejected interlopers.} Noting that the condition for the
recoil velocity experienced by a BH interloper to be larger than $v_{\rm esc }$ is 
$a\leq a_3=a_{\rm ej}/q_3^2$, then
the total number of BHs  ejected 
by a hard binary 
can be easily shown to be
\begin{equation}\label{inter}
N_{\rm 3}=\int^{a_{\rm m}}_{a_{\rm 3}}{1\over{\epsilon -1}}{da\over a}=
{1\over{1-\epsilon}}\ln \left(
{a_{\rm 3}\over a_{\rm m}}\right)\ ,
\end{equation}
where $\epsilon =
1/(1+\delta)$. { Eq.\ (\ref{inter})
shows that not only more massive clusters eject less binaries,
but also the number of ejected interlopers decreases significantly.}

Our numerical approach is similar to the method described in 
\citet{2016ApJ...831..187A}, but with at least one important difference.  \citet{2016ApJ...831..187A} calculated the binary hardening rate from 
the cluster {\it core} density which was set to be a constant
free parameter in the models,
and other calculations in 
the literature  followed a similar approach \citep[e.g.,][]{Miller2002,2009ApJ...692..917M,2018arXiv180901164C}.
Here,
we have used H\'{e}non's principle to relate
the hardening rate of the binaries to the evolving {\it global} properties
of their host cluster.

\subsection{Results}\label{resu}
In Fig.\ \ref{fig3} we show the mass of the merging binaries produced in four example cases
of cluster evolution.
We consider two cluster models with
the same initial mass, but different 
densities. For each model
we run two  realisations: one where
the spin parameter of the BHs is sampled from 
a uniform distribution in the range $[0,1)$, and the other where 
the BHs have no spin initially.
In the model with the higher density (lower panel),
 due to the shorter dynamical friction timescale, the BHs
 segregate more rapidly to the cluster core and start to merge earlier during
 the cluster evolution. Thus, it is the value of $\tau_{\rm df}$ for the most massive 
 BHs in the cluster that determines  the time
of the first merger (i.e., $100\,\rm Myr$ in the bottom panel,
 and $1\,\rm Gyr$ in the upper panel).
Higher densities also correspond to a higher binary hardening rate and retention 
fraction. For $\rho_{\rm h,0}=10^6\, M_\odot\rm\ pc^{-3}$
and initial spins in the range 0 to 1
($\chi=0$), we find that $1873$ ($1719$) mergers are produced in 
the first $1$\,Gyr, and of the merger remnants produced
over this time, $617$ (1400) are retained in the cluster.
For $\rho_{\rm h,0}=10^4 M_\odot\rm\ pc^{-3}$, $959$ (915) mergers are produced over $\sim 10\,\rm Gyrs$, and only 
$51$ (459) of the remnants are retained. 
Thus, higher densities lead to a higher retention fraction and faster evolution, and therefore to a larger probability that substantial growth 
occurs by the end of the run.
Accordingly,  while in 
the low density examples the merging 
binaries have masses $\mathcal{M} \lesssim 100\,M_\odot$,
 for
$\rho_{\rm h,0}=10^6 \,M_\odot\rm\ pc^{-3}$
a BH with $\mathcal{M}\sim 10^4\, M\odot$ is formed in  the first few Gyrs. 
As the BH grows, 
the mass ratio $q$ gets further away from unity and 
its dimensionless spin magnitude tends to decrease,
leading  to progressively lower GW kicks.
This keeps the growing BH safe in the cluster after
the first few mergers.

Fig. \ref{fig-maxmass} shows the
mass of the heaviest BH formed over 13\,Gyr of evolution,  $\mathcal{M}_{\rm max}$, as
a function of cluster properties.
For each value of initial mass and density, we evolved thirty random realisations
and show in the figure the mass of the
largest BH formed in each run. 
From Fig. \ref{fig-maxmass} we see  
that in sufficiently dense and massive clusters ($M_{\rm cl}\gtrsim 10^7 M_\odot$; 
$\rho_{\rm h,0}\gtrsim 10^5 M_\odot\rm pc^{-3}$),  BHs mergers can lead to substantial growth and produce massive seeds. Fig. \ref{fig-maxmass} shows
that for the range of stellar densities we considered $\rho_{\rm h,0}\le 10^7 M_\odot\rm pc^{-3}$,  only
clusters with total mass larger than several $10^6M_\odot$ produce massive seeds.

In Fig. \ref{fig-vesc} we plot $\mathcal{M}_{\rm max}$ as
a function of initial cluster escape velocity.
There is a clear transition near a cluster escape velocity of 
$v_{\rm esc,0}= 300\,\rm km\,s^{-1}$. Clusters with an initial escape velocity larger than this value are able to grow BHs with masses  above $1000\,M_\odot$ in less than a Hubble time.
Assumptions about metallicity and initial spins have little or no effect on this conclusion. 
{ The transition near $v_{\rm esc,0}= 300\,\rm km\,s^{-1}$ is explained mainly from the fact
that the mean value of the GW kick velocity distribution is  around $300\,\rm km\,s^{-1}$
for large spins  \citep[see Fig. 1 in][]{2016MNRAS.458.3075A}.}

For lower escape velocities, i.e.,
$v_{\rm esc}< 300\,\rm km\,s^{-1}$,
one finds that assumptions about spins and metallicity
affect the results in important ways (see Fig. \ref{zoom}).
In these  clusters, any growing BH is eventually ejected
 after a few mergers, so that the maximum mass that can be built though hierarchical mergers
 depends significantly  on the initial mass function of the BHs
 and their initial spins.
If the BHs have initially high spins, the  mass of the largest BH formed is about
twice the high mass end of the  BH progenitor population. Thus,
its value depends on metallicity.
For $Z=0.1Z_\odot$, 
$\mathcal{M}_{\rm max}\approx 100\,M_\odot$; for $Z=Z_\odot$, $\mathcal{M}_{\rm max}\approx 50\,M_\odot$.
If the BHs are formed with no spin, some fraction of the merger remnants will be retained due to the lowered GW recoil. This second generation BHs, however, are  formed with a large spin $\chi\simeq 0.7$, and they will be most likely ejected if they undergo a second merger.
It follows that the mass of the largest
 BH that can be formed 
 through repeated mergers
 in this case is
about {\it four} times
 the mass of the 
 largest BH that is produced 
by stellar evolution.
Accordingly, for $Z=0.1Z_\odot$ and non spinning BHs, our models do not typically  produce
BHs more massive than about $150 M_\odot$ (see Fig. \ref{zoom}).

\begin{figure}
\centering
\includegraphics[width=3.2in,angle=0.]{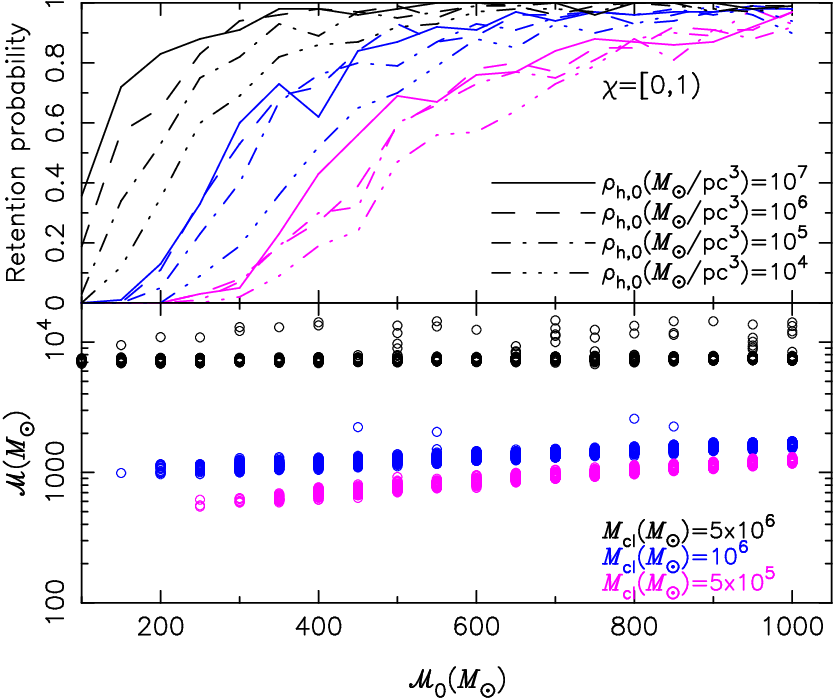}
\includegraphics[width=3.2in,angle=0.]{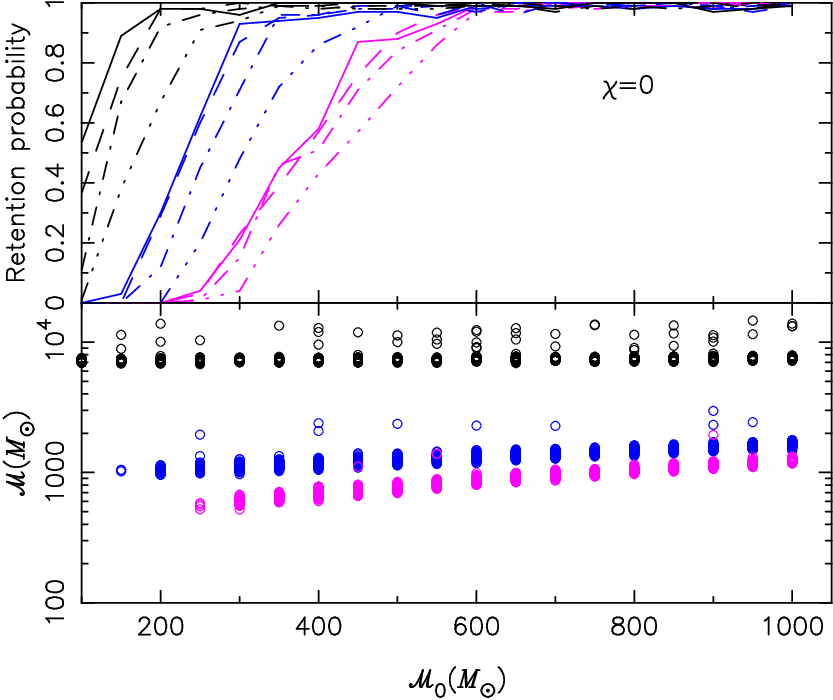}
 \caption{The upper panel in each figure gives the probability
 that a BH of initial mass $
\mathcal{M}_0$
 will still be inside its parent cluster after 13 Gyr of evolution. Cluster initial parameters
 are indicated; we set
 $Z=0.1Z_\odot$, and
 vary $
\mathcal{M}_0$ in the range $[100;1000]M_\odot$. 
For each cluster mass and density we simulated 100 models.
The  lower panels show
 the mass of the retained BH seeds at the end of the integration. 
 Upper figure is for initial BH spin magnitudes distributed
 in the range $\chi=[0,1)$; in the lower figure instead, the BHs have zero spin initially.
}
 \label{seed}
 \end{figure}

{
Our results appear to be in good agreement with those
of \citet{2018PhRvL.120o1101R}.
These authors 
investigated the formation of second generation BHs in their Monte Carlo  models of 
globular clusters. Thus  the comparison we make here only concerns systems similar to globular clusters ($v_{\rm esc,0} \lesssim 100\,\kms$). 
\citet{2018PhRvL.120o1101R} found  BH masses as high as $\simeq 150 M_\odot$,  but only get to $80M_\odot$ once the spins were increased
(see their Fig. 2).
}

{ \citet{2016MNRAS.458.3075A}
 predicted that BH binaries  observed by ground-based detectors should have similar masses, low spin magnitudes and  zero eccentricities, regardless of how they have formed. 
Thus, since current detectors only  observe a certain
subset of all mergers  they cannot distinguish formation mechanisms. The same authors later showed that a decihertz observatory could instead distinguish formation mechanisms as the eccentricity imprint from the formation channels is measurable at lower frequencies \citep{2017ApJ...842L...2C}.
We agree with these authors that the bulk of the BH merger populations produced by field and dynamical
channels should have similar properties in
the LIGO/Virgo frequency range, and that repeated mergers are unlikely to occur in any cluster due to the large
escape velocities required.
However, our models also show that the detection
of a   BH binary with total mass in the range $\simeq 100-200M_\odot$
will immediately  imply dynamical formation through hierarchical mergers in a nuclear  star cluster.  
Field formation (from both binaries and triple stars) and dynamical formation in globular clusters cannot produce
BH binaries with total masses in this range.
}

Finally, in Fig. \ref{fig-fraction} we present predictions for
the BH growth timescale (bottom panels), and the fraction of our models in which
 a BH of a given mass is formed after 13\,Gyr
 of evolution (upper panels).
 The probability that substantial growth  will occur in a cluster model
 appears to be a strong function of the cluster escape velocity.
The formation probability is  one for $v_{\rm esc,0} \gtrsim 300\,\kms$, and it  drops  sharply to zero for escape velocities smaller than this value.
The timescale over which the growth of the seed
occurs is most sensitive  to the cluster density.
We find that our cluster models with the highest densities 
(i.e., $\rho_{\rm h,0}\ge 10^6 M_\odot\rm pc^{-3}$),
are able to grow a
 BH with mass $\mathcal{M} \gtrsim 10^3\,M_\odot$ in a time $\lesssim \rm 1\,Gyr$, while for $\rho_{\rm h,0}= 10^5\, M_\odot\rm pc^{-3}$ it takes a time $\sim 10\,\rm Gyr$  to grow a comparable massive seed.

 An approximate value
 of the  minimum cluster  density   that is needed in order for 
 mass growth 
 to be able to occur
is obtained by requiring  that 
 the following two conditions are met:
 (i) $v_{\rm esc,0}>300\,\kms$, and
  (ii) the BH dynamical friction  timescale is shorter than a Hubble time, i.e.
 $t_{\rm df}(r_{\rm h})<13\,\rm Gyr$.
 Using Eq.(\ref{vescF}) and Eq.(\ref{dfr})
we find that conditions (i) and (ii) cannot be 
contemporarily satisfied for ${\rho}_{\rm h,0}\lesssim
10^4\,M_\odot{\rm pc^{-3}}
\left(m/ 20\,M_\odot\right)^{-1}
$, and therefore any BH mass growth should be suppressed in clusters with densities lower than this value.
This conclusion
appears to be in  agreement with the results shown in 
Fig. \ref{fig-fraction} where we see that for
 ${\rho}_{\rm h,0}=
10^5\,M_\odot{\rm pc^{-3}}$ it takes a
time $\sim 10\,\rm Gyr$  to grow any massive seed.
For densities smaller than this, 
any BH mass growth is suppressed.

The main take away from the analysis presented in this section  is that  in clusters with initial escape velocity  $\gtrsim 300\,\kms$ and  averaged  density
$\gtrsim 10^5\,M_\odot\rm pc^{-3}$,  repeated BH mergers should lead to the formation of
BHs  with mass above $10^3 M_\odot$  in less than a Hubble time.

\subsection{Clusters with a primordial massive seed}\label{primord}
In our models, the
initial mass function of the BHs has
a natural upper mass limit of about
$40M_\odot$. This limit is set by 
the pulsational pair instability supernovae \citep[e.g.,][]{2007Natur.450..390W,2014ApJ...792...28C},
and pair instability supernovae \citep{1967PhRvL..18..379B}.
 If the mass of the Helium core  is $\gtrsim 30M_\odot$, the formation of electron-positron pairs makes oxygen/silicon burn explosively. 
Hydrodynamical simulations  show that if the
helium core mass is 
$M_{\rm He}\lesssim64M_\odot$ the star experiences several pulses that enhance mass loss before the star forms a compact remnant, while
in the range 
$64\lesssim M_{\rm He}\lesssim135M_\odot$ the oxygen/silicon
ignition releases enough energy to disrupt the entire star \citep{2017ApJ...836..244W}.
For 
$M_{\rm He}\gtrsim 135 M_\odot$, however, the star is expected  to avoid
the pair instability and directly collapse to a massive remnant. 
\cite{2017MNRAS.470.4739S} show that
very massive stars with initial mass
$\gtrsim 200M_\odot$ and $Z<0.05Z_\odot$
form  BHs  with mass above $\gtrsim 200M_\odot$ via direct collapse.
This means that, although rarely, some  clusters
might form an initial massive seed well before BH mergers can play any important role. \cite{2018MNRAS.478.2461G} show that stars above $300M_\odot$ can also form via stellar collisions in the first few Myrs of evolution of a dense globular cluster.

After its formation, a  massive seed will inevitably  merge with other
BHs in the core, and following each merger it will receive
a GW recoil kick.
Although such kicks will be
 somewhat reduced
by the smaller mass ratios involved (see Eq.\ \ref{eqn:kick}),
they can still be sufficiently large to eject the seed from the 
cluster \citep{2008ApJ...681.1431M,2008ApJ...686..829H,2013A&A...557A.135K,2018ApJ...856...92F,2018ApJ...867..119F}.
Here we calculate the probability for this to happen.
We consider cluster models
with various masses and
densities, in which we introduced an initial
BH with mass in the range
$100 \leq \mathcal{M}_0\leq 1000 M_\odot$. For each
seed mass we integrated
100 cluster models with
a given density and mass, and set
$Z=0.1Z_\odot$.
We evolve these models
for 13 Gyrs, and plot 
in Fig.\ \ref{seed} 
the fraction of systems
in which the seed is
still in the cluster 
after
this time  (upper panels).
In the lower panels of
Fig.\ \ref{seed}, we plot
the final mass of the  seeds.

 Fig.\ \ref{seed} 
shows that whether 
a seed  will be 
kept  inside
its parent cluster or not
depends on the cluster mass.
For clusters with mass $\sim 10^6M_\odot$,
an initial seed
mass of $M_{0}\gtrsim 200M_\odot$
is needed for the retention 
probability to be significant.
The lower panels in Fig.\ \ref{seed} 
show that
the mass
of the growing seeds at the end of the evolution
is essentially independent of the
initial seed mass, and it is 
almost exclusively determined
by the cluster mass, 
in agreement with Eq.\ (\ref{Mmax3}) and our discussion in 
Section\ \ref{12rec}.
On average, the value of the final BH seed mass was also found to be   independent  of the initial cluster density.

\begin{figure}
\centering
\includegraphics[width=3.2in,angle=0.]{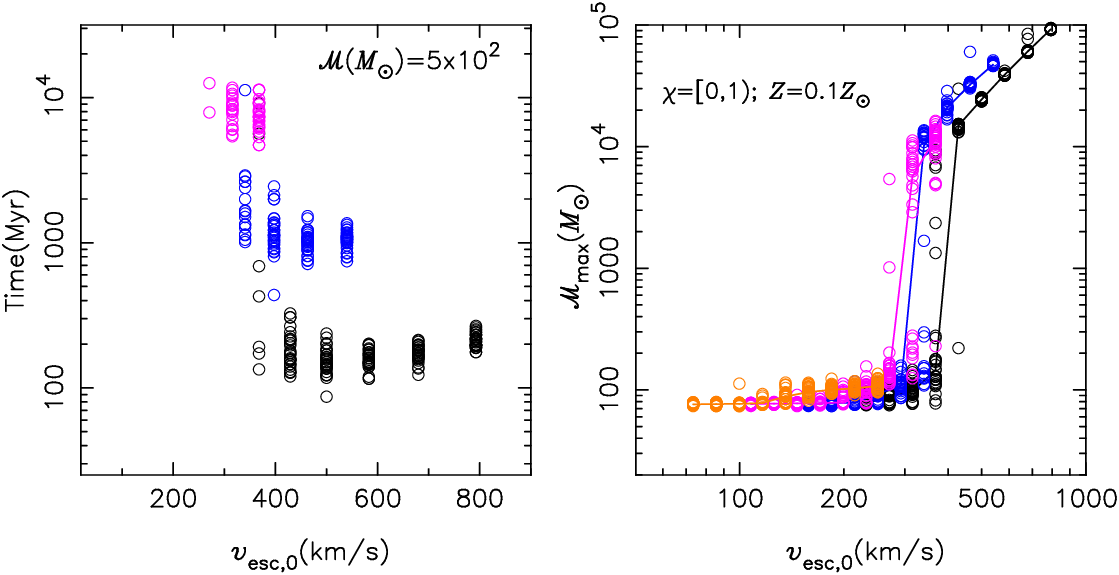}
 \caption{
The left panel shows the timescale needed to grow  a seed with mass above 
$5\times 10^2M_\odot$, and the right panel gives the
maximum mass of the binary BHs formed as a function of the cluster
initial escape velocity. 
Unlike our other models, in these simulations we have included the 
time delay due to the wandering of the BH seed off the cluster centre
caused by three-body superelastic scatterings.
A comparison to Fig. \ref{fig-vesc} (top-left panel) and Fig. \ref{fig-fraction} (bottom-left panel) shows that this effect
does not represent a limiting factor for
the growth of a massive seed. 
}
 \label{test}
 \end{figure}

\subsection{Approximations  and assumptions}\label{ass}
{ Our models are based on a number of assumptions 
which we now discuss and justify in more detail. 

We have assumed that the dynamical interactions only occur between BHs.
This is reasonable  because for
an evolved cluster, the densities of BHs near the core are expected
to be much larger than the densities of stars.
Hence, BHs will  dominate the interactions \citep[e.g.,][]{2013MNRAS.432.2779B,Morscher2015}.
Moreover, because
close three-body interactions pair the BHs with the highest mass
and tens of three body encounters 
are required before merger, then
mergers will be primarily
between BHs when they are present  \citep[e.g.,][]{Sigurdsson1993,2004ApJ...616..221G,
2009ApJ...692..917M}.
If the number of BHs in the core is low,  then they could exchange more frequently in
 mass-transferring binaries with main sequence, giant, and white dwarf companions 
\citep[e.g.,][]{2018ApJ...852...29K},
which  can lead to an observable electromagnetic signature \citep[e.g.,][]{Strader2012}.
It should be also noted  that in the most relevant case in which  dynamical ejections 
can be ignored (i.e., $v_{\rm esc}\gg v_{\rm 2-1}$),
then the binary hardening rate and merger time are only determined
by the cluster properties and are independent  of the interloper mass (see also  Section\ \ref{1rec}).





Another simplification we made  is to neglect the relativistic corrections of the orbits
during  binary single-interactions.
The most important corrections are the 2.5pN terms, which
for $\lesssim 5\%$ of the systems could lead
to a faster merger than predicted by our models.
For a detailed discussion of these effects see
 \citet{2006ApJ...640..156G} and \cite{2018PhRvD..97j3014S}.

Finally, we have set the binary merger time equal to $\tau_{\rm merge}=\tau_{\rm bin}+\tau_{\rm GW}$.
But the merger time will be exactly given by $\tau_{\rm merge}$
only if the binary remains confined near the core region so that
the  rate of binary-single interactions does  not drop significantly below the value given by Eq.\ (\ref{eq:edotbin}).
This might not be a good
approximation at early stages 
when the seed is ``light''
and can be kicked  
off the centre due to three body interactions, causing the hardening rate to go down or to nearly stop
\citep{2016ApJ...819...70M,2018MNRAS.475.1574D}. 
However,  the energy balance argument implies that 
the overall merger rate of binaries should  be hardly affected by
this -- if the binary spends a significant amount of time outside the core region
a new binary must form in the core in order to generate the required heating rate, leaving
the BH merger rate unaffected. 
Furthermore, 
 the displacement due to the GW recoils (which are  included in our
 models above)  and the resulting time delay are typically much larger than those due to superelastic
 three-body encounters
 which therefore should have a secondary effect on the growth timescale of the seed.

To demonstrate the latter point, we ran a new suit of simulations
in which  the time 
the binary spends outside the core was included in the calculation of its merger time.
Following the numerical procedure   described in \citet{2018MNRAS.481.5445S}, we divide  the dynamical evolution of a binary in
 $N_{\rm 2-1}$ isolated binary-single interactions that lead 
to a stepwise decrease in the binary semi-major axis until $a\approx a_{\rm m}$.
Following  the  $i^{\rm th}$ interaction,
 the binary semi-major
axis decreases from $a^i$ to $\epsilon a^i$.
The
  binary centre of mass receives
a kick $v_{2-1}$ and it is deposited at a distance 
$r_{\rm in}(v_{2-1})$ from the centre.
Given the new binary orbit, we
 compute the time for 
 the next interaction to 
occur as 
$\tau_{\rm df}^{i}(r_{\rm in})+\tau^{i}_{\rm enc}$, where  
$\tau^{i}_{\rm enc}\simeq 0.2{Gm_1m_2\over 2\epsilon a^i} \dot{E}_{\rm bin}^{-1}$
and  $\dot{E}_{\rm bin}$ is computed at time $t+\tau^{i}_{\rm df}$.
The  total lifetime of the binary
is then $\tau_{\rm merge}=\tau_{\rm GW}$+$\sum_i^{N_{\rm 2-1}}(\tau_{\rm df}^{i}+\tau^{i}_{\rm enc})$.
We consider  models with $Z=0.1Z_\odot$ and initial BH spins sampled uniformly in the
range $\chi = 
\left[0, 1\right)$. The results from these new models
are presented in Fig. \ref{test}.
A comparison  with the results from the simulations described above (see Fig. \ref{fig-vesc}  and Fig. \ref{fig-fraction}) 
confirms  that accounting for the delay time due to dynamical kicks
 has a small effect on
 the growth timescale  and final mass of the seed.
  
}

\section{Astrophysical implications}\label{astro}

\subsection{Globular clusters}
A number of studies have argued that ground based GW observations will be able 
to address whether intermediate mass BHs (IMBHs; generally defined  as BHs with mass in the range $\sim 100-10^5 M_\odot$)   can form in globular clusters via hierarchical mergers
of stellar seed BHs \citep[e.g.,][]{2002ApJ...576..894M,2004ApJ...616..221G,2015MNRAS.454.3150G,2018PhRvD..97l3003K,2018ApJ...858L...8C}. 
Our dynamical models
show, however, that
 in clusters with parameters consistent with present-day globular clusters ($M_{\rm cl}\lesssim 10^7M_\odot$,
 $\rho_{\rm h}\lesssim 10^5 M_\odot/\rm pc^{3}$), it is not
possible to form BHs above $\approx 100M_\odot$ through mergers of smaller BHs (see Fig. \ref{nuclear-clusters}), because their initial escape velocities are too low ($\lesssim 300\,\kms$). 
This  result  appears to be rather
insensitive to the assumptions about dynamics and  about the initial distribution of spins and masses of the BHs.
On the other hand, globular clusters could  have formed  a
massive {BH via mergers} in the past if
their {escape velocities}  were initially much larger than  today; for example, a cluster
mass of $M_{\rm cl}\lesssim 10^6M_\odot$ ($10^5M_\odot$)
would require an initial half-mass radius  $r_{\rm h,0}\lesssim 0.1$pc (0.01pc). 
As shown in Fig.\ \ref{nuclear-clusters}
such extremely compact clusters
will then expand within a Hubble time to
become the larger clusters we observe today.

We conclude that if future observations will show that IMBHs exist at the centre of globular clusters,
then the scenario presented here
could be a plausible explanation to  their origin
and would imply that these systems form
at high redshift with extremely large densities 
($\gtrsim 10^7M_\odot\rm pc^{-3}$).
Other scenarios, invoking
runaway collisions of massive stars and/or the collapse of a very massive star ($\gtrsim 200M_\odot$)
could also lead to the formation of IMBHs in globular clusters (e.g., 
\cite{2002ApJ...576..899P},
\cite{2004ApJ...604..632G},
 \cite{2004MNRAS.355..413P}, \cite{Zwart2004a},
\cite{2006MNRAS.368..141F}, \cite{2016MNRAS.459.3432M}, \cite{2018MNRAS.478.2461G}, and Section \ref{primord}). {A GW detection may be the most promising route to search for IMBHs in globular clusters, because a convincing detection with other methods remains elusive \citep{ 2010ApJ...710.1032A,2013ApJ...769..107L, 2018MNRAS.473.4832G,2018ApJ...862...16T,2018PhRvD..98f3018A}.}

\subsection{Nuclear star clusters}

Nuclear star clusters (NCs) are the most massive and densest clusters observed in the local universe
\citep{1997AJ....114.2366C,2004AJ....127..105B,2006ApJS..165...57C}.
\citet{2009ApJ...692..917M} and \citet{2016ApJ...831..187A} 
argued that in most NCs,
due to their large escape velocities, dynamical ejections
can be reasonably ignored.
Under these conditions, 
 Eq.\ (\ref{rate})  gives the total BH merger rate produced 
 by a cluster of a given mass and radius, and can be used to determine 
the total merger rate per volume produced locally by NCs:
\begin{align}\label{rateV1}
\Gamma_{\rm tot}\simeq 
{{dN_{gx}\over dV}  
}  f_{\rm BH} {1\over N_{\rm NC}}\sum_{i=1}^{N_{\rm NC}} \
 \Gamma_{{\rm m},i}(M_{\rm cl},r_{\rm h})
\end{align}
where  $\Gamma_{\rm m}$
is the BH merger rate given by Eq.\ (\ref{rate}),
 $N_{\rm NC}$ is the total number of clusters
in the sample considered,
${dN_{gx}/dV}$ is the number density of nucleated galaxies in the local universe, and $f_{\rm BH}$ is the fraction of these galaxies without a massive BH.
Because the value of $f_{\rm BH}$ is poorly constrained by observations, we leave it as a parameter
in our calculations below, but note that 
values of $f_{\rm BH}\gtrsim 0.5$ are 
consistent with the results
of   galaxy formation models
\citep{2015ApJ...812...72A}, while
 observations prefer marginally smaller values
 $f_{\rm BH}\sim 0.2$ \citep{2008ApJ...678..116S,2018ApJ...858..118N}.
Most probably $f_{\rm BH}$ should also be a function of
galaxy mass and morphology,
which we ignore here.
We assume that the BHs have all the same mass which  we take to be $10M_\odot$.
We set  ${dN_{gx}/dV}=0.01\rm Mpc^{-3}$ \citep{2005ApJ...620..564C} and assume an order of unity occupation 
fraction of NCs in galaxies, \citep{2012ApJS..203....5T,2014MNRAS.441.3570G}.

\begin{figure}
\centering
\includegraphics[width=3.in,angle=0.]{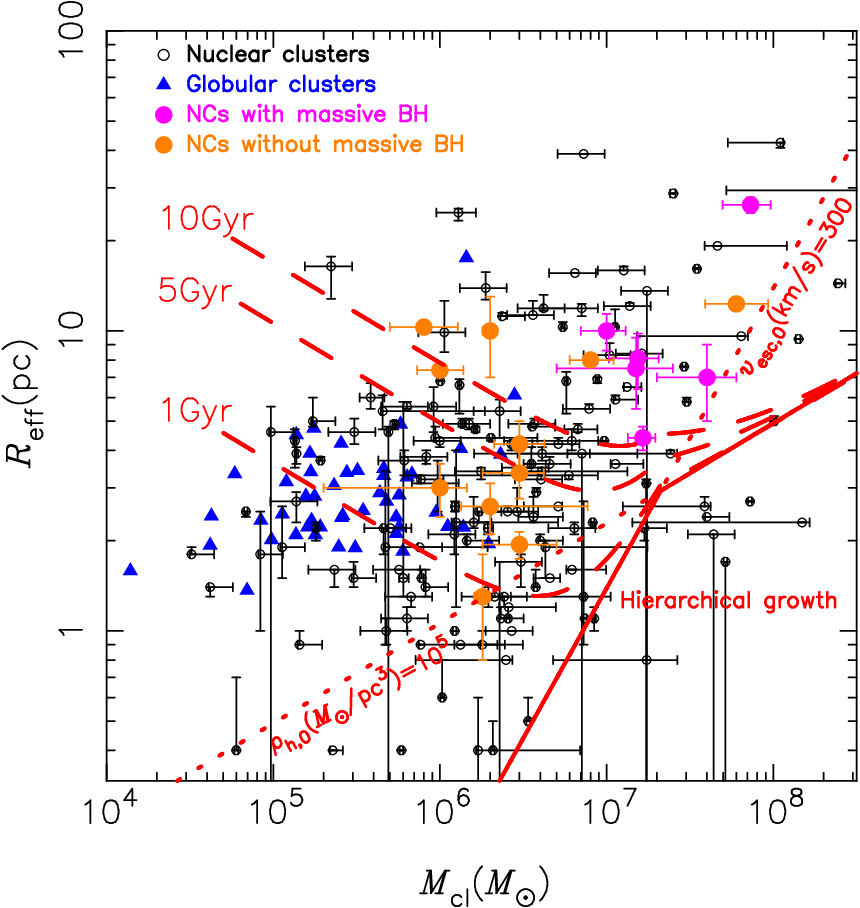}
 \caption{
 Effective radii and masses of NCs
 from \citet{2016MNRAS.457.2122G}, and globular
 clusters from \citet{2008MNRAS.389.1924F}.
 Below the solid-red line our models predict that 
an IMBH will grow in less than a Hubble time. The dashed-red lines show the effective radius of  clusters  with initial density $\rho_{\rm h,0}=10^5 M_\odot\rm pc^{-3}$ 
 and escape velocity $v_{\rm esc,0}=300\rm km\ s^{-1}$
after evolving them for 1, 5 and 10 Gyr (from bottom to top line).   If we assume mass follows light, then $r_{\rm h} \simeq 4/3 R_{\rm eff}$. 
Purple symbols are systems for which a central massive BH  has been measured.
Orange symbols are NCs for which there is only an upper limit to the central BH mass.
See text for additional details.
  }
 \label{nuclear-clusters}
 \end{figure}

{ 
Because the merger rate is determined at any given time
 by the {\it current} cluster properties
through  Eq. (\ref{rate}), it is not affected by the details of how the cluster has evolved and formed
at earlier times nor on the galaxy morphology and mass.
Thus, in order to compute  the local merger rate of BH binaries  produced by 
a population of NCs we only need to know their  present-day
values of $r_{\rm h}$ and $M_{\rm cl}$, assuming that these clusters
contain BHs.} 
We compute $\Gamma_{{\rm m}}(M_{\rm cl},r_{\rm h})$
for  the  $N_{\rm NC}=151$ clusters in the observational sample
of \citet{2016MNRAS.457.2122G} with
a  well determined mass and effective radius (see Fig. \ref{nuclear-clusters}).
Then, from  Eq.\ (\ref{rateV1}) we find $\Gamma_{\rm tot}\approx  6f_{\rm BH} \rm  Gpc^{-3} yr^{-1}$. 
Although necessarily approximated, this calculation shows that NCs (without a massive BH) contribute significantly to the overall merger rate of binary BHs in the local universe
\citep[see also][for a similar result]{2016ApJ...831..187A}.
Including the contribution from NCs hosting a massive BH could
increase these rates significantly \citep{OLeary2009,2012ApJ...757...27A,2016ApJ...828...77V,2017ApJ...846..146P,2018ApJ...865....2H}.

We showed that clusters with 
escape velocity $v_{\rm esc,0}\gtrsim 300\,\kms$ and 
initial densities $\gtrsim 10^5\,\mspc$, can  grow a BH with mass well above $\approx 100M_\odot$
in less than a Hubble time.
Such higher densities and   escape velocities
 lead to higher  retention fractions
and merger rates, favouring 
rapid growth.
Now, we turn to the  question of
whether such  extreme conditions  are met 
in NCs.
Fig. \ref{nuclear-clusters} gives
the total mass and half-light radius (or effective radius) of
the NCs in the sample of
\citet{2016MNRAS.457.2122G}.
This sample comprises the NCs in spheroid-dominated
galaxies from \cite{2006ApJS..165...57C} and \cite{2012ApJS..203....5T},  and disk dominated galaxies from \cite{2009MNRAS.396.1075G} and \cite{2014MNRAS.441.3570G}.
The range of plausible parameters 
for which, according to our previous analysis, hierarchical growth is likely to occur are indicated 
by the box in the bottom-right corner of the figure.
This region of parameter space
is defined as that where  both conditions $v_{\rm esc,0}\gtrsim 300\,\kms$ and 
 $\rho_{\rm h,0}\gtrsim 10^5\,\mspc$
 are met.
Remarkably, we find that 
17  out of the 151 of the NCs in the sample
have structural parameters that are consistent with being inside
this region of parameter space.
Hence a  fraction
$f_{\rm NC} \approx 0.1$ 
of the present-day NCs meet the conditions 
required for fast hierarchical BH mergers to occur within their cores.
{ If present day NCs with the required properties exist today,
then they must have been more
 abundant in the early Universe when densities where higher.
This motivates searches for IMBHs with mass modelling approaches in present day NCs (that do not longer have the required 
densities).}

 \begin{figure}
\centering
\includegraphics[width=2.65in,angle=0.]{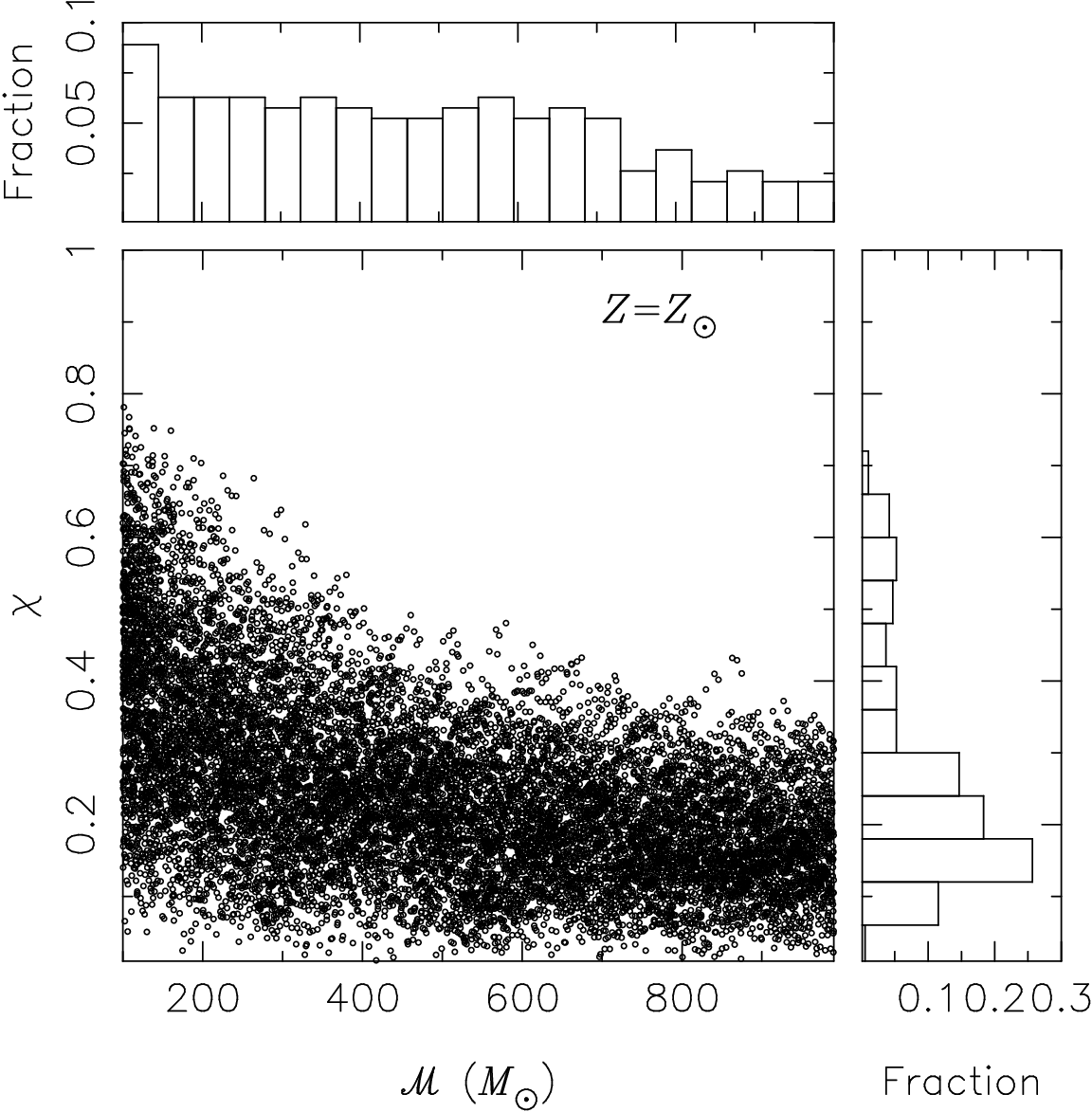}
\includegraphics[width=2.65in,angle=0.]{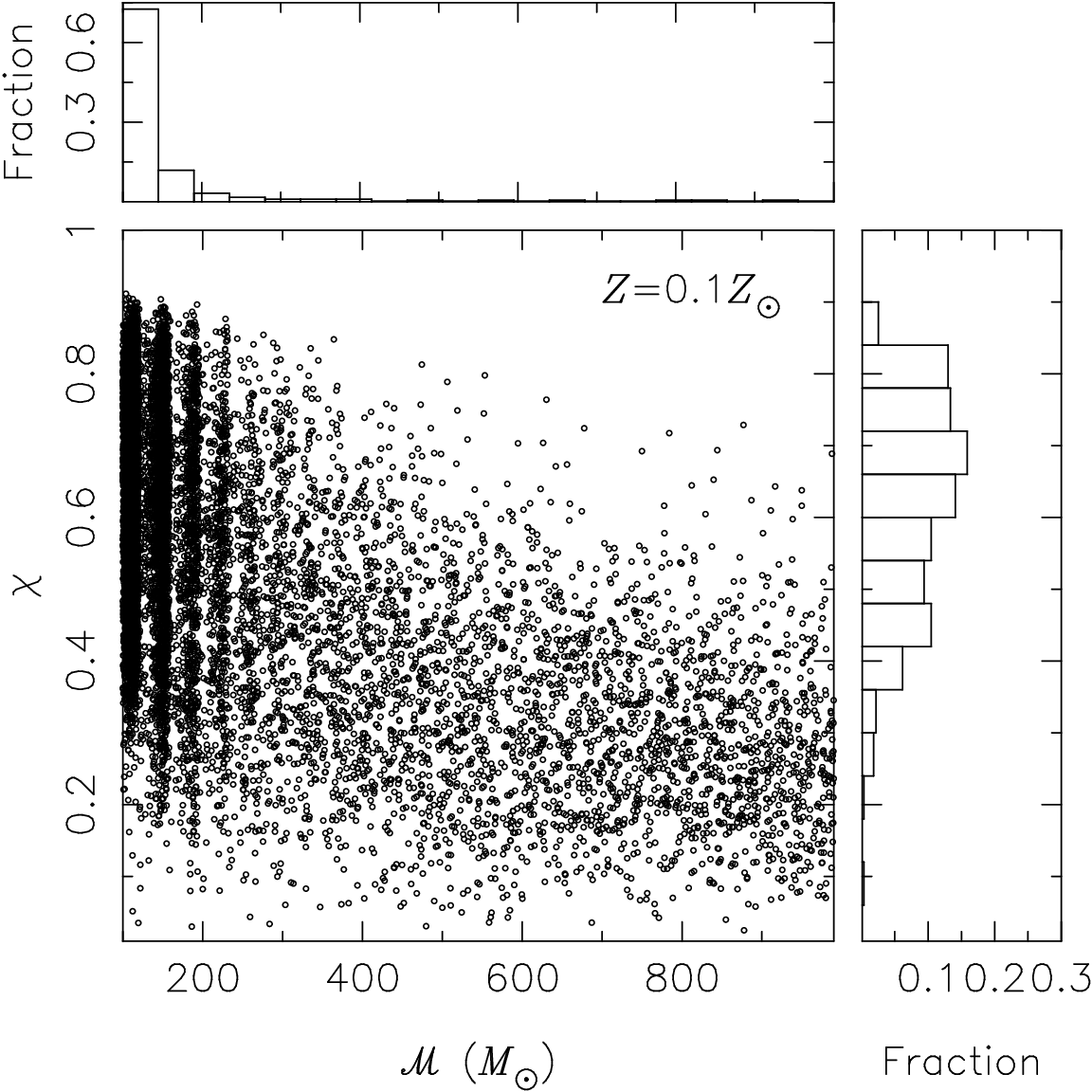}
 \caption{
 Intrinsic spin and mass distributions for the 
 merging binaries with  mass
$\mathcal{M}>100M_\odot$ formed  in our cluster models (cluster radius and mass 
were uniformly distributed inside the region where hierarchical growth is possible, i.e., 
$v_{\rm esc,0}\ge 300\rm km\ s^{-1}$ and  $ \rho_{\rm h,0}\ge 10^5M_\odot\rm\ pc^{-3}$). 
Upper panel is for Solar metallicities,
lower panel is for $Z=0.1Z\odot$.
 Because the spin parameter
 of the BHs tends to decrease with the number of mergers, a correlation exists 
 between  mass and spin of the BHs, with the heavier BHs having the smaller spins.
  }
 \label{fig-chimass}
 \end{figure}

If several BHs with masses  $>100 M_\odot$ are formed, 
these will provide  strong sources of GWs detectable by Advanced LIGO/Virgo through their merger and/or ringdown
\citep[e.g.,][]{2002ApJ...581..438M,2015PhRvL.115n1101V}. 
We compute the corresponding merger rate
per  volume  
as
\begin{align}\label{rateV}
\Gamma_{\rm V}^{\rm }\simeq 
 {{dN_{gx}\over dV}  
} f_{\rm BH} f_{\rm NC} {1\over N_{\rm NC}}\sum_{i=1}^{N_{\rm NC}} \
 \Gamma_{{\rm m},i}(M_{\rm cl},r_{\rm h})
\end{align}
where  $f_{\rm BH}$ 
should be intended here as the fraction of NCs in the relevant region of parameter space that do not have a massive BH, and $\Gamma_{{\rm m},i}(M_{\rm cl},r_{\rm h})$ the merger
rate of BHs in the mass range $\mathcal{M} =[10^2,10^3]M_\odot$.
{ We compute the sum in Eq.\ (\ref{rateV})
by evolving a synthetic population of 
 $N_{\rm NC}=100$ cluster models
 with  mass and  radius distributed
uniformly below the solid-red line in 
Fig.\ \ref{nuclear-clusters}, and impose
the conditions  $M_{\rm cl}\leq 10^8\,M_\odot$, and  $\rho_{\rm h,0}\le 10^7\,\mspc$.} We
evolved these models until
a time $t_{\rm f}=(2^{3/2}-1){2\over3}\tau_{\rm rh,0}/\zeta$, after which time  the clusters  have expanded significantly and   $r_{\rm h}(t_{\rm f})=
2 r_{\rm h,0}$.
The basic assumption   is that there exists 
a constant number
of clusters
in the high density region of parameter space
where hierarchical growth 
occurs -- this requires replenishment 
of the clusters  on a time-scale
$\sim t_{\rm f}$ (i.e., several times the
cluster relaxation timescale). 
Under this assumption, the merger rate of large BHs for each simulated cluster is simply $\Gamma_{\rm m}\simeq N_{\rm m}/t_{\rm f}$, with $N_{\rm m}$ the  number of BH binary mergers in the
mass range
$\mathcal{M} =[10^2,10^3]M_\odot$.
We evolved two sets of models
with $Z=Z_\odot$ and
$0.1Z_\odot$, and found that 
they  produce a
BH merger rate per  volume of  $\Gamma_{\rm V}\simeq 0.04f_{\rm BH}\, \rm Gpc^{-3} yr^{-1}$
and $0.06f_{\rm BH}\, \rm Gpc^{-3} yr^{-1}$, respectively.
This merger rate  
can be compared with existing limits 
derived from the non-detection of IMBHs
by Advanced LIGO. For 
 $100-20\,M_\odot$  binaries, \citet{2017PhRvD..96b2001A}
reported an upper limit to the merger rate of $\lesssim 10\,\rm Gpc^{-3} yr^{-1}$
at $90\%$ confidence, broadly consistent with our findings.

Finally, from our models we  compute the detection rate
of large BHs by Advanced LIGO/Virgo as
\begin{align}\label{rateO}
\Gamma_{\rm obs}\simeq 
{{dN_{gx}\over dV}  
} f_{\rm BH} f_{\rm NC}{1\over N_{\rm NC}}\sum_{i=1}^{N_{\rm NC}}{1\over{t_{{\rm f},i}}} \sum_{j=1}^{N_{\rm m}}\
  {4\pi\over 3}D_{j}^3
\end{align}
where $D_{j}$ is the luminosity  distance out to which 
the ring-down phase of the $j^{\rm th}$ event can be detected with a 
S/N of at least 10. 
We computed $D_{j}$
using Eq.\ (A17) in \citet{1998PhRvD..57.4535F},
and ignore cosmological corrections
which  for current interferometers
are small \citep[e.g.,][]{2002ApJ...581..438M}.
For $Z=Z_\odot$, our models 
give 
 a detection rate
$\Gamma_{\rm LIGO}\simeq  0.3f_{\rm BH}\, \rm yr^{-1}$ for events with total mass $>100\,M_\odot$.
Even a small  contribution from
lower metallicity clusters increases
substantially these rates.
Taking $Z=0.1Z_\odot$, 
 our models 
give 
$\Gamma_{\rm LIGO}\simeq 3f_{\rm BH}\,\rm yr^{-1}$.
Thus, if $\sim 10\%$ of  the local high density/mass NCs  have  such metallicities (or lower), then 
the total detection rate will be increased to
$\Gamma_{\rm LIGO}\simeq  0.6f_{\rm BH}\, \rm yr^{-1}$.

 BHs that are formed from previous mergers admit a unique mass/spin correlation that could allow our model to be tested in the near future. This is demonstrated in
Fig.\ \ref{fig-chimass} where we show the mass
and spin distribution of the large BHs
formed in the NC models. 
After the first few mergers,
we expect $\chi_{\rm }\approx 0.7$ \citep{2017PhRvD..95l4046G,2017ApJ...840L..24F}.
But, 
as the random walk process of  accretion 
of smaller BHs progresses, $\chi_{\rm }$  decreases
so that the larger BHs will also have  the smaller spins
\citep[e.g.,][]{2008ApJ...681.1431M}.
For $\mathcal{M}\gtrsim 200M_\odot$, we find that
spins are typically confined in the range $\lesssim 0.6$.
The final spin distribution is sensitive to
the initial mass function of the BHs, 
and therefore to metallicity.
Because low metallcity stars produce larger BHs, fewer mergers are needed to reach a certain mass, resulting in
a larger spin parameter at a given $\mathcal{M}$.

\subsection{Formation of massive black hole seeds}
The end-product of our merger scenario is
the production of  IMBHs.
Subsequently to their formation, these seeds 
could then grow by swallowing  stars and/or through standard Eddington-limited 
accretion of gas to become the massive BHs we observe today.
A weak test to this idea 
is provided in  Fig.~\ref{nuclear-clusters}. Here we show the
population of NCs with strong evidence for a massive BH (purple symbols) and NCs with only an upper limit 
to the mass of a putative central BH (orange symbols).
We find a sharp transition   from clusters without to
with evidence of a central BH at  mass larger than 
several $ 10^6M_\odot$.
This transition is consistent with our previous
analysis showing
that the formation of BH seeds should be favoured 
for clusters with masses larger than this value 
(see Section \ref{resu}).
For our scenario to be valid, however, we require not only that
the cluster masses are large, but also that the clusters are initially sufficiently
dense to allow for rapid growth.
Hence, we compare the present-day clusters' radii to the
radii of  cluster models with  $\rho_{\rm h,0}= 10^5\,\mspc$ 
and $v_{\rm esc,0}=300\,\kms$
that we evolved  using Eq.~(\ref{rex}) for 1, 5 and 10\,Gyr (dashed-red lines).
All, but one, of the observed systems
have radii that are  near, although somewhat larger than, the 
predicted radii after 10\,Gyr of evolution. 
{The radii as a function of mass are similar to the results in \citet{2010MNRAS.408L..16G}, albeit  somewhat larger. This is because we fixed $\psi=5$, where in reality the clusters at 10\,Gyr have narrower mass functions and hence $\psi\approx2$.}
{However, the initial radii could have been much smaller, because}
the Universe was much denser in the past, by a factor $(1+z)^3$.
This suggests that significant expansion of the cluster, and BH growth, might have occurred
in these systems over the Hubble time.

Finally, 
we note that episodic star formation and accretion of star clusters 
can lead to morphological and structural transformation of the nuclei which is difficult to
address with  our simplified models \citep[e.g.,][]{2013ApJ...763...62A,2015ApJ...812...72A}.  A  self-consistent 
assessment of our  scenario for massive BH seed formation will therefore require 
to model the cluster formation and evolution in a cosmological set.  
Binaries with masses between
a hundred and a few thousand Solar masses have frequencies that will  make them detectable
by space-borne gravitational-wave observatories such as LISA \citep{2007CQGra..24R.113A,2018LRR....21....4A}.
Third-generation detectors such as the proposed Einstein Telescope 
\citep{2010CQGra..27h4007P}
and the Cosmic Explorer \citep{2017CQGra..34d4001A}
will be able to probe GWs in a frequency range reaching down to $\sim $1Hz and detect BH mergers at high redshift.
This will provide 
direct constraints on the proposed mechanism, which are independent of the uncertain 
assumptions about the dynamics and evolutionary history of the clusters.

\bigskip

 FA acknowledges support from an STFC Rutherford fellowship (ST/P00492X/1) from the Science and Technology Facilities Council.
 MG acknowledges financial support from the Royal Society (University Research Fellowship) and the European Research Council (ERC-StG-335936, CLUSTERS).

\bibliographystyle{mnras}

\end{document}